\documentclass[traditabstract]{aa} 

\usepackage{amssymb}
\usepackage{amsmath}
\usepackage{hyperref}
\usepackage{graphicx}
\usepackage{natbib}
\usepackage{booktabs}
\usepackage{caption}
\usepackage{placeins}
\usepackage[toc,page]{appendix} 
\usepackage{gensymb}

\usepackage{comment}
\usepackage{threeparttable}
\usepackage{multicol}
\usepackage{tabularx}
\usepackage{lscape}
\def\GLEE{\textsc{Glee}}

\usepackage{color}
\usepackage{xcolor}

\begin{document}

   \title{The missing quasar image in the gravitationally lensed quasar HE0230$-$2130: Implications for the cored lens mass distribution and dark satellites}

  \titlerunning{Modeling HE0230$-$2130}

  \author{S. Ertl\inst{1}\inst{,2}
                  \and
                  S. Schuldt\inst{3,4}
                  \and
                  S. H. Suyu\inst{1}\inst{,2}\inst{,5}
                  \and
                        P. L. Schechter \inst{6}
                                          \and
                        A. Halkola \inst{7} \and
                        J. Wagner\inst{8}
}

        \institute{Max-Planck-Institut f{\"u}r Astrophysik, Karl-Schwarzschild Stra{\ss}e 1, 85748 Garching, Germany\\
             e-mail: \href{mailto:ertlseb@mpa-garching.mpg.de}{\tt ertlseb@mpa-garching.mpg.de}
         \and
             Technical University of Munich, TUM School of Natural Sciences, Department of Physics,  James-Franck-Stra{\ss}e 1, 85748 Garching, Germany
        \and 
        Dipartimento di Fisica, Universit\`a  degli Studi di Milano, via Celoria 16, I-20133 Milano, Italy
        \and 
        INAF - IASF Milano, via A. Corti 12, I-20133 Milano, Italy
          \and
             Academia Sinica Institute of Astronomy and Astrophysics (ASIAA), 11F of ASMAB, No.1, Section 4, Roosevelt Road, Taipei 10617, Taiwan
             \and
            MIT Kavli Institute for Astrophysics and Space Research, Cambridge, MA 02139, USA
            \and 
            Pyörrekuja 5 A, FI-04300 Tuusula, Finland
            \and
            Bahamas Advanced Study Institute and Conferences,
            4A Ocean Heights, Hill View Circle, Stella Maris, Long Island, The Bahamas
        }

   \date{Received --; accepted --}

  \abstract{Strongly lensed systems with peculiar configurations allow us to probe the local properties of the deflecting lens mass while simultaneously testing general profile assumptions. The quasar HE0230$-$2130 is lensed by two galaxies at similar redshifts ($\Delta z \sim 0.003$) into four observed images. Using modeled quasar positions from fitting the brightness of the quasar images in ground-based imaging data from the Magellan telescope, we find that lens-mass models where each of these two galaxies is parametrized with a singular power-law (PL) profile predict five quasar images.  One of the predicted images is unobserved despite it being distinctively offset from the lensing galaxies and likely bright enough to be observable. This missing image gives rise to new opportunities to study the mass distribution of these galaxies. To interpret the quad configuration of the system, we tested 12 different profile assumptions with the aim of obtaining lens-mass models that correctly predict only four observed images. We tested the effects of adopting: cored profiles for the lensing galaxies; external shear; and additional profiles to represent a dark matter clump. We find that half of our model classes can produce the correct image multiplicity. By comparing the Bayesian evidence of different model parametrizations, we favor two model classes: (i) one that incorporates two singular PL profiles for the lensing galaxies and a cored isothermal sphere in the region of the previously predicted fifth image (rNIS profile), and (ii) one with a bigger lensing galaxy parametrized by a singular PL profile and the smaller galaxy by a cored PL profile with external shear.
 We
estimated the mass of the rNIS clump for each candidate model of our final Markov chain Monte Carlo sample, and find that only 2\% are in the range of $10^6 M_{\odot} \leq M_{\rm rNIS}\leq 10^9 M_{\odot}$, which is the predicted mass range of dark matter subhalos in cold dark matter simulations, or the mass of dark-matter-dominated and low-surface-brightness galaxies. We therefore favor the models with a cored mass distribution for the lens galaxy close to the predicted fifth image. Our study further demonstrates that lensed quasar images are sensitive to the dark matter structure in the gravitational lens. We are able to describe this exotic lensing configuration with relatively simple models, which demonstrates the power of strong lensing for studying galaxies and lens substructure.}


   \maketitle
%
\section{Introduction}
\label{sec:intro}

Strong gravitational lensing (SL) describes the relativistic effect where a massive object along our line of sight distorts the light of a background source into multiple arcs. These objects are called lenses and can be single galaxies or even groups and clusters of galaxies. In the case of lensed quasars, we observe multiple images that are especially prominent as the quasars are extremely bright compared to their host galaxies. Because the light rays interact gravitationally with both baryonic and dark matter (DM), SL allows us to study the local properties of the deflector \citep{Kochanek1991,Rusin2000,Cohn2001,Suyu2009,Sonnenfeld2011,Wagner2019}, as well as the total mass distribution or the DM fraction \citep{Bolton2006, Barnabe2011,Gavazzi2012,Sonnenfeld2015}. 

An important part of these studies is to obtain a mass model of the deflector by choosing a mass parametrization and optimizing its parameters to fit the data. For galaxy lenses, the most common parametrization is that of a power-law (PL) profile. The PL class of lens models arises for any ensemble of objects that form a structure via gravity as the dominating interaction \citep{Wagner2020}. SL studies on early-type galaxies (ETGs), such as that of the Sloan Lens ACS (SLACS) Survey \citep{Bolton2006}, have provided important insights. One example is the average slope of the total mass density, which is found to be well described by a PL profile with a small intrinsic scatter around $\gamma\sim2$ \citep{Auger2010, Barnabe2011, Shajib2021}. The three-dimensional mass density is $\rho(r) \propto r^{-\gamma}$, where $\gamma$ is the PL slope. 

Insights into other lens-mass parameters of ETGs were gained by \citealt{Bolton2008}, who modeled 63 SLACS lenses with singular isothermal ellipsoids. The values for the external shear of each lens range from 0 to 0.27 (median of 0.05), and the axis ratios range from 0.51 to 0.97 with a median of 0.79 \citep{GomerTh2020}. 
Independent studies of the lens environment determined similar values for the external shear \citep{Wong2011}. Similarly, the distribution of axis ratios from SLACS is consistent with values from studies of nearby elliptical galaxies \citep{Ryden1992}. 

The SLACS sample probes a low-redshift range up to $z\sim0.5$ with a median redshift of $z\sim0.2$ \citep{Gavazzi2012}.
Another lens sample is from the Strong Lensing Legacy Survey (SL2S), which spans a range of $z=0.2-0.9$ with a median redshift of $z=0.5$, making it a good high-redshift complement to the SLACS survey, as both surveys probe galaxies with similar sizes and velocity distributions \citep{Jackson2010}. One key result from analyzing the SL2S lenses is that the total mass density slopes of individual galaxies do not appear to evolve over time \citep{Sonnenfeld2015} while the average slope of the population of lens galaxies becomes steeper over time \citep[e.g.,][]{Bolton2012, Sahu2023}.

Early-type galaxies are of very high interest in astrophysics as their formation and evolution are still unclear. They are characterized by their old stellar populations, red colors, small amount of cold gas and dust, and lack of spiral arms \citep{Cappellari2016}.
The current picture is that ETGs are the result of a hierarchical merging scenario: big structures form through the merger of smaller structures \citep{Toomre1972, Kauffmann1993,White1991,Cole2000}. During a merger, the central regions of the galaxies can be disrupted by the interaction of two supermassive black holes (SMBHs), which remove stars from the central regions, leading to the formation of a core \citep{Faber1997,Milosavljevic2002}. Another requirement for core formation is the lack of cold gas, which, if present, would lead to a nuclear star burst that would fill the depleted region in the center. Studies show that the mass of the SMBH in the center correlates with the mass absent from the nucleus \citep{Graham2005,Kormendy2009,Rusli2013,Dullo2014}. The cores formed in this way in the center of massive ellipticals typically have sizes of between tens of parsecs and a few hundred parsecs (e.g., \citealt{Dullo2014}). Cores larger than 1 kpc are rare. The largest observed cores are of the order of 3 kpc; for example, that observed in the massive brightest cluster galaxy of Abell 2261 \citep{Postman2012,Nasim2021}.

The core is evident in the surface brightness profile of ETGs, with a shallow inner profile and a steeper outer profile \citep{King1966,Lauer1995}. Insights into the structural parameters and formation histories of ETGs therefore provide a strong test for the $\Lambda$CDM cosmological standard model. Recently, observations by the James Webb Space Telescope in the JADES \citep{Bunker2019} and CEERS \citep{Finkelstein2022} surveys revealed tensions with the $\Lambda$CDM model. The modeled stellar masses in these galaxies at large redshifts (z\textgreater10) are very high and should be much lower in these young galaxies according to current models \citep{Labbe2023}. The agreement with cosmological simulations is better \citep{Mccaffrey2023}, but nevertheless galaxy evolution remains a matter of high interest and studies of ETGs play a crucial role.
\\

Strong gravitational lensing can also be used to detect DM subhalos or low-mass dark galaxies (i.e., galaxies with a high DM content and a low surface brightness). Structure formation through hierarchical merging is incomplete: the inner parts of DM halos can survive and remain as subhalos within their new host. CDM simulations  confirm these predictions and show that these subhalos have cusped inner density profiles and an upper mass limit of around $10^6 M_{\odot} - 10^9 M_{\odot}$ \citep{Dalal2001,Diemand2008}. This predicted mass range is similar to those of faint, DM-dominated dwarf satellites \citep{Simon2007,Vegetti2012}. 

There are several approaches to using SL for substructure detection. One approach involves the study of flux-ratio anomalies, where the observed flux ratios of the multiple source images differ from those predicted by the smooth-mass model \citep{Mao1997,Metcalf2001,Xu2014,Nierenberg2017,Gilman2019, Hsueh2020}. The flux ratios are sensitive to perturbations in the lensing potential and are therefore an indication of small-scale structure in the halo of the lensing galaxy, such as dark matter substructure. 

Another approach to the detection of DM substructure uses a method called gravitational imaging, where a best-fitting smooth-lens model is compared to the data and substructures are detectable through anomalies in the surface brightness distribution of the lensed arc \citep{Falco1999,Koopmans2005,Vegetti2010,Vegetti2014,Nierenberg2014,Ritondale2019, Despali2020, Bayer2023a, Bayer2023b}.\\

Here, we present a modeling analysis of the quadruply lensed quasar HE0230$-$2130. The configuration of this system is peculiar as there are two lensing galaxies that are possibly part of a galaxy group and may be dynamically interacting to some extent. One of the four quasar images is located between them. For a lensing configuration like this, we would expect five quasar images, and so our goal in this work is to find model parametrizations that can explain a missing quasar image. The lensed quasar PS J0630$-$1201 \citep{Ostrovski2018,Lemon2018} has a very similar configuration to HE0230$-$2130, but a fifth image is observed.
Double galaxy lenses like HE0230$-$2130 and PS J0630$-$1201 are uncommon and can cause exotic lens configurations with complex critical and caustic curves \citep{Orban2008}. Understanding these systems might reveal better probes of the underlying lens-mass distribution, or, in the case of high image magnification, may provide ways to resolve even more distant objects than before. Other quasars lensed by binary galaxies are, for example, 2M1310$-$1714 \citep{Lucey2018}, DES J0408$-$5354 \citep{Lin2017,Agnello2017}, and B1608+656 \citep{Myers1995,Suyu2009}.

We use ground-based imaging data from the Magellan telescope and use the multiple modeled image positions as constraints for the lens-mass distribution. There are HST data available but there is no suitable guide star in the field. A drizzled image of the available unguided and short exposures leads to a noisy image, and therefore the Magellan data are currently the optimal data for our analysis. Approaching this system naively by describing each of the two lensing galaxies with a PL mass distribution results in models that do not fit our observed data: as expected, they always predict one additional, distinctively offset and bright quasar image that is not observed. 

To find models that are in agreement with the observations, we analyzed 12 different parametrizations with varying degrees of complexity: singular or cored PL profiles and an additional singular or cored isothermal, spherical mass clump close to the location of the previously predicted fifth image. Each model class is analyzed with and without the presence of external shear. By quantitatively comparing these different classes of models, we can probe possible reasons for the missing fifth image and place constraints on the lens-mass distributions of HE0230-2130.\\

The outline of the paper is as follows: in Sect.~\ref{sec:observations} we describe the observations of this lensing system, in particular the imaging data used for the modeling.  In Sect.~\ref{sec:microlensing}, we discuss whether microlensing, dust extinction, or quasar variability could be responsible for the missing image. We then describe our modeling methods and assumptions in Sect.~\ref{sec:modeling}. The modeling results are presented and discussed in Sect.~\ref{sec:results}. In Sect.~\ref{sec:summary}, we summarize our findings.

Throughout the paper, we adopt a flat $\Lambda$CDM cosmology with $H_0=70 \rm{\, km\, s^{-1}\, Mpc^{-1}}$ and $\Omega_{\rm M} = 1 - \Omega_{\Lambda}=0.3$. Parameter estimates are given by the median of the one-dimensional marginalized posterior probability density functions, and the quoted uncertainties show the 16$^{\rm th}$ and 84$^{\rm th}$ percentiles (corresponding to a 68\% credible interval).

\section{Observations of HE0230$-$2130}
\label{sec:observations}

\begin{figure}
\centering
        \includegraphics[trim={0 0 0 2cm},clip,scale=0.5]{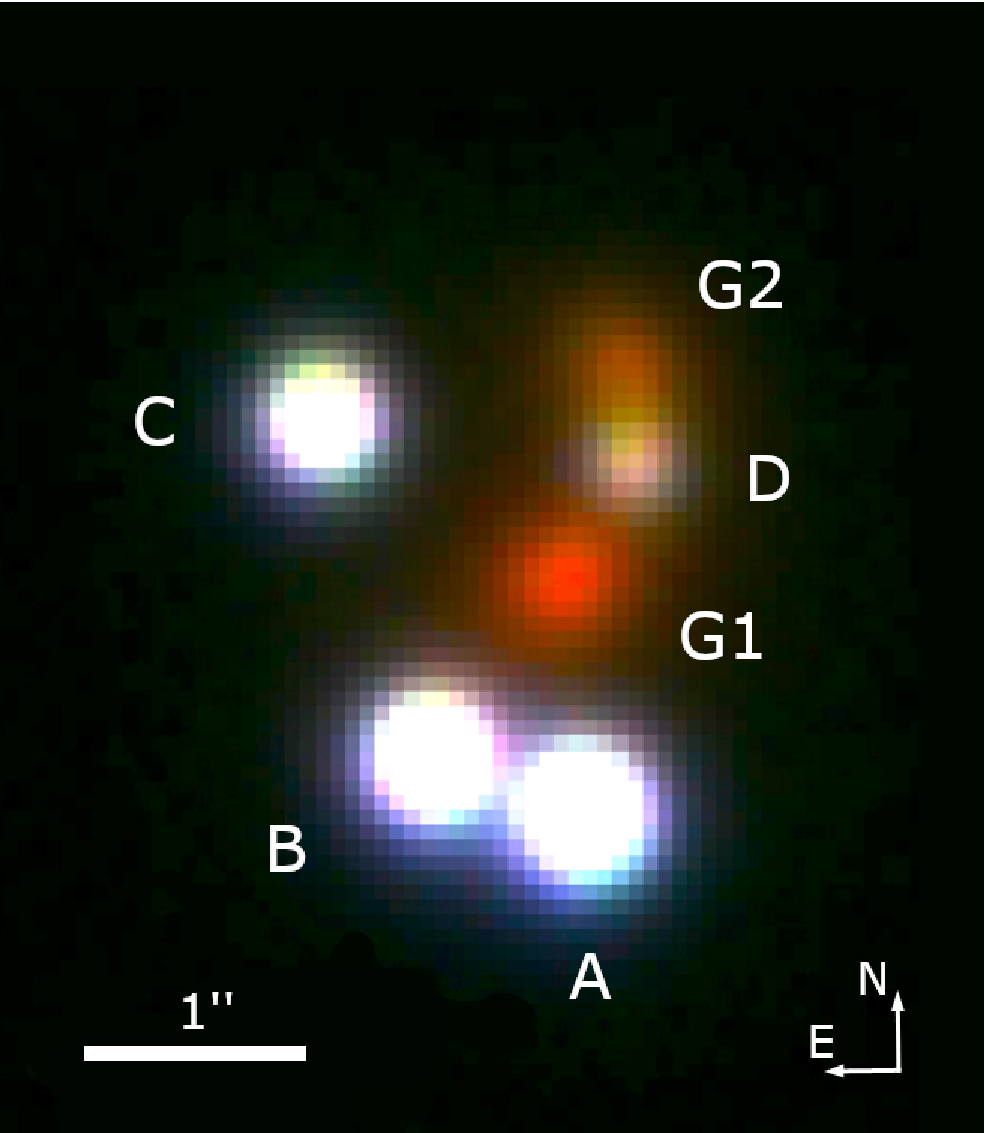}
        \caption{Color image of HE0230$-$2130 created with Magellan imaging data in the $g$-, $r$-, and $i$-band. Image courtesy of Scott Burles.}
        \label{fig:colorimg}
\end{figure}

This quadruply imaged quasar at redshift $z_s=2.162$ was discovered by \citet{Wisotzki1999} as part of the Hamburg/ESO survey (\citealt{Wisotzki1996}). It is located at (right ascension (RA), declination (dec))=(38.13829, -21.29046). In 2005, images of this system were taken at the Magellan Clay Telescope with MagIC, which we use here in the modeling. Figure~\ref{fig:colorimg} shows a color image, which is a combination of three 400s exposures in the $g$- ,$r$-, and $i$-filter of the Magellan data with a pixel size of 0.069\arcsec.  The seeing was approximately 0.41\arcsec, 0.35\arcsec, and 0.35\arcsec, respectively.

The brightest lensing galaxy (G1) is located between the multiple lensed quasar images, while the other, fainter galaxy (G2) is slightly offset to the north of image D. The maximum image separation is 2.15\arcsec. \citet{Faure2004} found a galaxy overdensity about 40\arcsec south-west of the lens and proposed that G1 and G2 are two members of a physically related galaxy group. This was later confirmed spectroscopically by \citet{Eigenbrod2006}, with measured redshifts of $z_{\rm G1}=0.523\pm0.001$ and $z_{\rm G2}=0.526\pm0.002$. The spectrum of G1 is a good match to that of an ETG. \citet{Anguita2008} obtained similar redshifts for G1 and G2, and found the redshift of a galaxy $\sim17\arcsec$ northwest of image A to be $z=0.518\pm0.002$. To understand whether or not G1 and G2 are potentially dynamically associated with each other we determined whether or not a typical peculiar velocity of cluster galaxies can account for the observed redshift difference of 0.003 \citep{Harrison1974}. We find that we need a peculiar velocity of around 1000 km/s, which is a typical value for cluster galaxies. Therefore, G1 and G2 seem to be dynamically associated and dynamical interaction is possible to some extent.

\section{The missing image}
\label{sec:microlensing}
In this section, we discuss some physical reasons that could suppress the fifth image (subsequently referred to as image E) using the predictions from a mass model with two singular PL profiles for the two lens galaxies. Given the predicted proximity of image E to lensing galaxy G2, we need to assess whether microlensing or dust extinction can have a significant effect on the flux of the image and make it unobservable. 

\subsection{Microlensing}
Microlensing is caused by compact objects, such as stars, in the lens galaxy, which can further magnify or demagnify the lensed quasar images. \citet{Weisenbach2021} estimated worst-case microlensing scenarios. Given the median convergence of $\kappa\sim0.8$ and assuming a stellar mass fraction at the location of image E of $\sim0.2,$ which is realistic for our system, the worst-case magnitude change for this saddle image is $\Delta_{--}\sim1.5$, which corresponds to a magnification ratio of roughly 0.25 between the worst-case demagnified image E and the image E unaffected by microlensing; this means that image E is demagnified through microlensing by a factor of 4  at most.

\begin{figure}
\centering
        \includegraphics[trim={0 0 0 0cm},clip,scale=0.5]{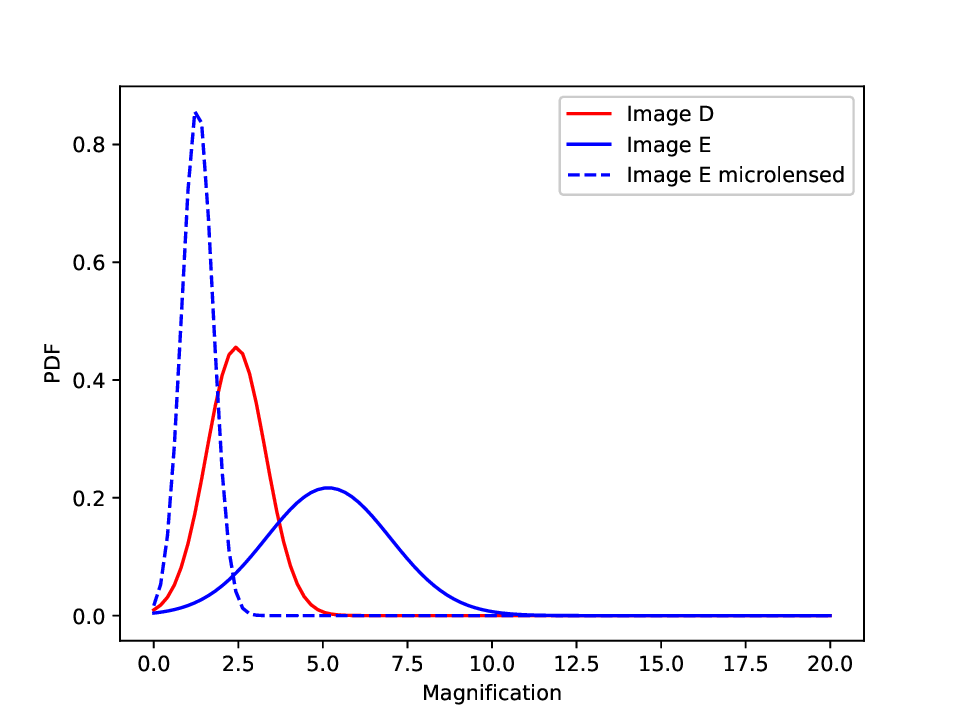}
        \caption{Distributions of predicted image magnifications for the observed image D (red, solid line), the model-predicted image E (blue, solid line), and the demagnified image E in a worst-case microlensing scenario (blue, dashed line). Even if image E were strongly demagnified by microlensing, it would still be observable.}
        \label{fig:mag_D_E}
\end{figure}

Figure~\ref{fig:mag_D_E} shows the model-predicted magnifications of image D (red, solid curve), image E (blue, solid curve), and image E in the case of a worst-case microlensing demagnification (blue, dashed curve). The median magnification of image E when unaffected by microlensing is $\sim4.8$. The median of the distribution of microlensed image E ($\sim 1.2$) is lower than the median magnification of image D ($\sim2.3$). Both of these values are above the lowest magnification of $\sim0.05$ for which an image would be observable in our Magellan image.
Therefore, microlensing is an unlikely explanation for the missing image.

\subsection{Dust extinction}
Dust in the lensing galaxy can also reduce the observed fluxes of lensed quasar images. \citet{Anguita2008} found that image D is somewhat affected by dust extinction by fitting an extinction law to the flux-ratio spectrum of image A/image D. To check for dust extinction with our Magellan data, we also modeled the lens and quasar light in the $g$-band and calculated the color excess compared to the $i$-band. We find no significant color excess compared to the other three quasar images. As image E has a similar distance to G2, we suspect that it is affected by dust to a similar extent to image D.  The effect of dust seems to be small, and so even the combined effect of microlensing and dust is not likely to be enough to completely extinguish image E.

\subsection{Intrinsic quasar variability}
Another possibility is that the natural variability of the quasar and the time delay between the lensed images have conspired to temporarily reduce the flux of image E below the detection threshold. 
The magnitudes of the quasar images changed by only about $\Delta m \sim 0.5$ from 2004 to 2017 \citep{Millon2020}. For the image to become unobservable, the magnitudes would need to increase by $\Delta m \sim5$, and so this is also not a possible explanation for the missing image.

\section{Modeling method}
\label{sec:modeling}
Having shown that the missing image cannot be explained by microlensing, dust, or variability, as described in the previous section, we looked for answers in modeling this system with different assumptions on the mass profiles. We model HE0230$-$2130 with the software Gravitational Lens Efficient Explorer \citep[\GLEE,][]{Suyu.2010,Suyu.2012} and its Bayesian sampling and optimization methods (simulated annealing and MCMC methods), which were automated by \citet{Ertl2023} and \citet{Schuldt2022} to minimize user interaction time. As these automated methods find the optimal sampling parameters and start new chains completely independently, we used them here to speed up the sampling of multiple different model classes.\\
\subsection{Light and mass parametrization}
To obtain accurate image positions and priors for the mass parameters of our models, we started by modeling the observed surface brightness of the quasar images and the two lensing galaxies. We selected the $i$-band of the Magellan data because the lensing galaxies and quasar images are the brightest in this band. 

The point-spread function (PSF) is constructed from the cutout (3.38\arcsec$\times$3.38\arcsec) of a single star in the field at (RA, dec)=(38.1548, $-$21.3166). There are three stars in the MagIC field of this observation, but we obtained the best results by choosing only this particular star for the PSF. As the star is not necessarily located at the center of the brightest pixel, we interpolated and shifted the cutout within fractions of a pixel to center the star. We subsampled the PSF by a factor of 3 because the pixel size of 0.069\arcsec\ is large relative to the full width at half maximum (FWHM) of the PSF (the FWHM of the PSF covers only a few pixels in the native data resolution). Thus, we avoided numerical inaccuracies caused by an undersampled PSF. The light of the quasar images is modeled by fitting this PSF with three parameters: the $x$- and $y$-centroid and the amplitude, which is the flux of the respective image.

We parametrized the light of each lensing galaxy with one S\'{e}rsic profile \citep{Sersic.1963} and the quasar images with a point source. Both were convolved with the PSF to fit to the observed image. The S\'{e}rsic profile is parametrized as
\begin{equation}
        I_{\rm S}(x,y) = A_{\rm S}\exp\Bigg[-k\Bigg\{\Bigg(\frac{\sqrt{(x-x_{\rm S})^2+\left( \frac{y-y_{\rm S} }{q_\text{S}} \right) ^2}}{r_{\rm eff}}\Bigg)^{\frac{1}{n_{\rm s}}}-1\Bigg\}\Bigg],
        \label{eq:sersic}
\end{equation}
where $A_{\rm S}$ is the amplitude, $x_{\rm S}$ and $y_{\rm S}$ are the centroid coordinates, $q_{\rm S}$ is the axis ratio, and $n_{\rm S}$ the S\'{e}rsic index. The constant $k$ normalizes $r_{\rm eff}$ such that $r_{\rm eff}$ is the half-light radius in the direction of the semi-major axis. The light distribution is oriented with a position angle $\phi_{\rm S}$ that is measured east of north (where $\phi_{\rm S}=0^\circ$ corresponds to the major axis being aligned with the northern direction).

In general, we parametrize the lens-mass distribution of G1 and G2 with a PL profile. The dimensionless surface mass density (or convergence) of a general PL profile is given by 
\begin{equation}
\label{eq:kappa_spemd}
\begin{split}
        \kappa_{\rm PL}(x, y) = \frac{3-\gamma}{1+q}\frac{\theta_{\rm E}^2}{(\theta_{\rm E}^2+r_{\rm c}^2)^{\frac{3-\gamma}{2}}-r_{c}^{3-\gamma}} \\
        \Big[(x-x_{\rm G})^2 + \frac{(y-y_{\rm G})^2}{q^2}+r_c^2\Big]^{\frac{1-\gamma}{2}},
\end{split}
\end{equation}
where $(x_{\rm G}, y_{\rm G})$ is the lens-mass centroid, $q$ is the axis ratio of the elliptical mass distribution, $\theta_{\rm E}$ is the Einstein radius, $r_{\rm c}$ is the core radius, and $\gamma$ is the radial profile slope. The mass distribution is oriented with a position angle of $\phi$. For $r_{\rm c}=0\arcsec$, this reduces to a singular PL elliptical mass distribution (\citealt{Barkana.1998}). The case with $\gamma=2$ and $q=1$ corresponds to an isothermal sphere profile. We refer to isothermal spheres with $r_{\rm c}=0$ and  $r_{\rm c}>0$ as a singular isothermal sphere (SIS) and  a non-singular isothermal sphere (NIS) profile, respectively. 
We can account for the tidal gravitational field of surrounding objects by adding an external shear component \citep{Keeton1997}. The shear strength is calculated as $\gamma_{\rm ext} = \sqrt{\gamma_{\rm ext,1}^2 + \gamma_{\rm ext,2}^2}$, with $\gamma_{\rm ext,1}$ and $\gamma_{\rm ext,2}$ as the components of the shear matrix. The lens potential for the external shear is parametrized by $\psi_{\rm ext} = \frac{1}{2}\gamma_{\rm ext}[\textrm{cos}(2\phi_{\rm ext})(x^2-y^2)+2\textrm{sin}(2\phi_{\rm ext})xy]$, where $\phi_{\rm ext}$ is the shear position angle. The position angle is measured east of north, which means that, for $\phi_{\rm ext}=0^\circ$, the images are stretched vertically, and for $\phi_{\rm ext}=90^\circ$ the images are stretched horizontally.

\subsection{Model classes}
\begin{figure}[h]
\centering
        \includegraphics[scale=0.6]{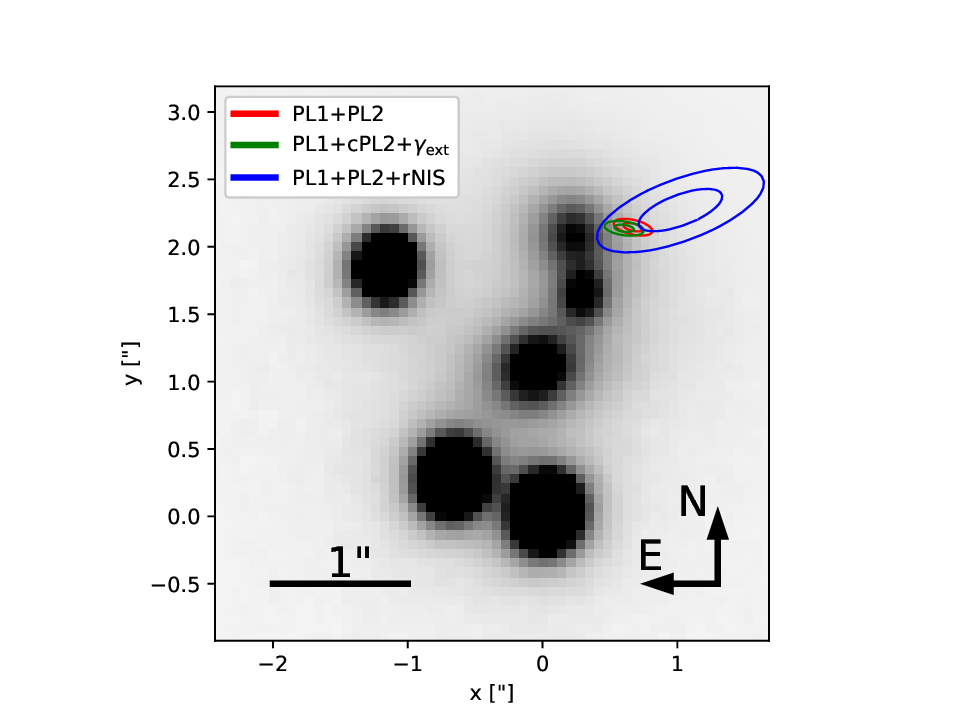}
        \caption{Modeled image of HE0230$-$2130. Plotted are the 1$\sigma$ and 2$\sigma$ contours of the predicted positions of the unobserved image E in the PL1 + PL2 (red), the PL1+cPL2+$\gamma_{ext}$ (green), and the PL1+PL2+rNIS (blue) model class. These predicted fifth quasar images would be above the detection threshold, but are not observed in the imaging data.}
        \label{fig:2pl_5thimg}
\end{figure}

After obtaining a lens and quasar light fit, we use only the four modeled quasar image positions as constraints for our models as there is no extended structure (lensed arc) visible in the data. In addition, we assume that G1 and G2 are located at the same redshift. From this lensing configuration, we would expect five images. The straightforward approach of modeling this system by parametrizing G1 and G2 with a PL profile and adding external shear always produces one additional, unobserved quasar image about 0.5\arcsec\ west of G2 (see Fig.~\ref{fig:2pl_5thimg}). Our goal is therefore to find models that can produce the observed number of quasar images of HE0230$-$2130. To this end, we set up 12 different model parametrizations, each with varying degree of complexity and physical motivation. G1 and G2 are each described either by a singular (PL) or cored PL (cPL) profile with or without external shear. We also consider models with a singular or cored isothermal, spherical mass clump in the region of the predicted image E (see Fig.~\ref{fig:2pl_5thimg}). We call these profiles ``restricted'' SIS/NIS (i.e., rSIS and rNIS). Table~\ref{tab:modelclasses} shows an overview and description of the multiple model classes. 
In all models, we fix the mass centroids of G1 and G2 to the modeled S\'{e}rsic centroid of the galaxies. While the axis ratio of the PLs of G1 and G2 are free to vary, we set a Gaussian prior on the position angle based on the light. The prior on the position of the rSIS and rNIS is Gaussian and centered on a typical position of image E. We choose a small $\sigma=0.14\arcsec\sim2$ pixels because the mass clumps tend to move away from the region of predicted image E even with a relatively narrow flat prior. The reason for this is that we do not penalize models that produce more than four images during the sampling. A detailed overview of the priors on the mass and shear parameters in each model is shown in Table~\ref{tab:priors}.

\begin{table*}[h!]
\caption{Overview of all 12 model classes.}
\centering
                \begin{tabular}{lp{.7\linewidth}}\toprule
                        Model name & Description \\ \midrule \midrule
                        \phantom{c }PL1 + \phantom{c}PL2 & G1\&G2: singular PL profiles\\ \rule{0pt}{2ex}
                        \phantom{c}PL1 + \phantom{c}PL2 + $\gamma_{\rm ext}$ &  G1\&G2: singular PL profiles, external shear\\ \rule{0pt}{2ex}
                        \phantom{c}PL1 + cPL2 & G1: singular PL profile, G2: cored PL profile \\ \rule{0pt}{2ex}
                        \phantom{c}PL1 + cPL2 + $\gamma_{\rm ext}$ & G1: singular PL profile, G2: cored PL profile, external shear \\ \rule{0pt}{2ex}
                        cPL1 + cPL2 & G1\&G2: cored PL profiles \\ \rule{0pt}{2ex}
                        cPL1 + cPL2 + $\gamma_{\rm ext}$ & G1\&G2: cored PL profiles, external shear \\ \rule{0pt}{2ex}
                        \phantom{c}PL1 + \phantom{c}PL2 + \phantom{r}SIS & G1\&G2: singular PL profiles, SIS\hspace{2px} without positional prior\\ \rule{0pt}{2ex}
                        \phantom{c}PL1 + \phantom{c}PL2 + \phantom{r}NIS & G1\&G2: singular PL profiles, NIS without positional prior\\ \rule{0pt}{2ex}
                        \phantom{c}PL1 + \phantom{c}PL2 + rSIS & G1\&G2: singular PL profiles, SIS\hspace{2px} with positional prior\\ \rule{0pt}{2ex}
                        \phantom{c}PL1 + \phantom{c}PL2 + rNIS & G1\&G2: singular PL profiles, NIS with positional prior \\ \rule{0pt}{2ex}
                        \phantom{c}PL1 + \phantom{c}PL2 + rSIS + $\gamma_{\rm ext}$ & G1\&G2: singular PL profiles,  SIS\hspace{2px} with positional prior, external shear \\ \rule{0pt}{2ex}
                        \phantom{c}PL1 + \phantom{c}PL2 + rNIS + $\gamma_{\rm ext}$ & G1\&G2: singular PL profiles,  NIS with positional prior, external shear \\
                        \bottomrule
                        \rule{0pt}{2ex}

                \end{tabular}
                \ \
        \caption*{\textbf{Notes: } The positional prior on the SIS and NIS models has a Gaussian distribution and is centred in the region where additional images were predicted.}
        \label{tab:modelclasses}
\end{table*}

\begin{table*}[h]
\caption{Model parameters and their priors.}
                \begin{tabular}{llp{3cm}p{2cm}p{6cm}} \toprule
                        Component & Parameter & Description & Prior & Prior range / value \\ \midrule \midrule
                        & $x_{\rm G}\ [\arcsec]$   & $x$-centroid & exact & fixed to best-fit light model \\ \rule{0pt}{2ex}
                         & $y_{\rm G}\ [\arcsec]$   & $y$-centroid  & exact & fixed to best-fit light model  \\ \rule{0pt}{2ex}
                        G1/G2 & $q$         & axis ratio & flat & [0.6, 1]\\ \rule{0pt}{2ex}
                        (PL/cPL)& $\phi$ [$^\circ$]   & position angle & Gaussian & centered on best-fit light model, $\sigma=10^{\circ}$\\ \rule{0pt}{2ex}
                        & $\theta_{\rm E}\ [\arcsec]$  & Einstein radius & flat & [0.1\arcsec, 1.2\arcsec]\\ \rule{0pt}{2ex}
                        & $r_{\rm c}\ [\arcsec]$  & core radius (cPL) & flat & [0\arcsec, 0.4\arcsec]\\\rule{0pt}{2ex}
                        & $\gamma$      & power-law index & Gaussian & center: 2.08, $\sigma=0.04$\\ \midrule
                        
                        & $x_{\rm sat}\ [\arcsec]$   & $x$-centroid & Gaussian & center: 0.73\arcsec, $\sigma=0.14\arcsec$   \\ \rule{0pt}{2ex}
                        satellite & $y_{\rm sat}\ [\arcsec]$   & $y$-centroid  & Gaussian & center: 2.17\arcsec, $\sigma=0.14\arcsec$  \\ \rule{0pt}{2ex}
                        (rSIS/rNIS) & $\theta_{\rm E, sat}\ [\arcsec]$  & Einstein radius & flat & [0.0001\arcsec, 1\arcsec]\\ \rule{0pt}{2ex}
                        & $r_{\rm c, sat}\ [\arcsec]$ & core radius (rNIS) & flat & [0\arcsec, 1\arcsec]\\ \midrule
                        
                        external& $\gamma_{\rm ext}$ & strength & flat & [0, 0.4]\\ \rule{0pt}{2ex}
                        shear& $\phi_{\rm ext}$ [$^\circ$] & position angle & flat & [$0^\circ, 180^\circ$]
             \\
             \bottomrule
             \end{tabular}
                \ \\ \ \\
        \caption*{\textbf{Notes: } The mass centroids of G1 and G2 are fixed to the light centroids of G1 and G2 from the best-fit light model, while the position angle of the mass has a Gaussian prior centered on the best-fit light model. The prior on the PL slope follows the SLACS lens sample \citep{Auger2010,Shajib2021}.}
        \label{tab:priors}
\end{table*}

\begin{figure*}[h!]
\centering
        \includegraphics[scale=0.16]{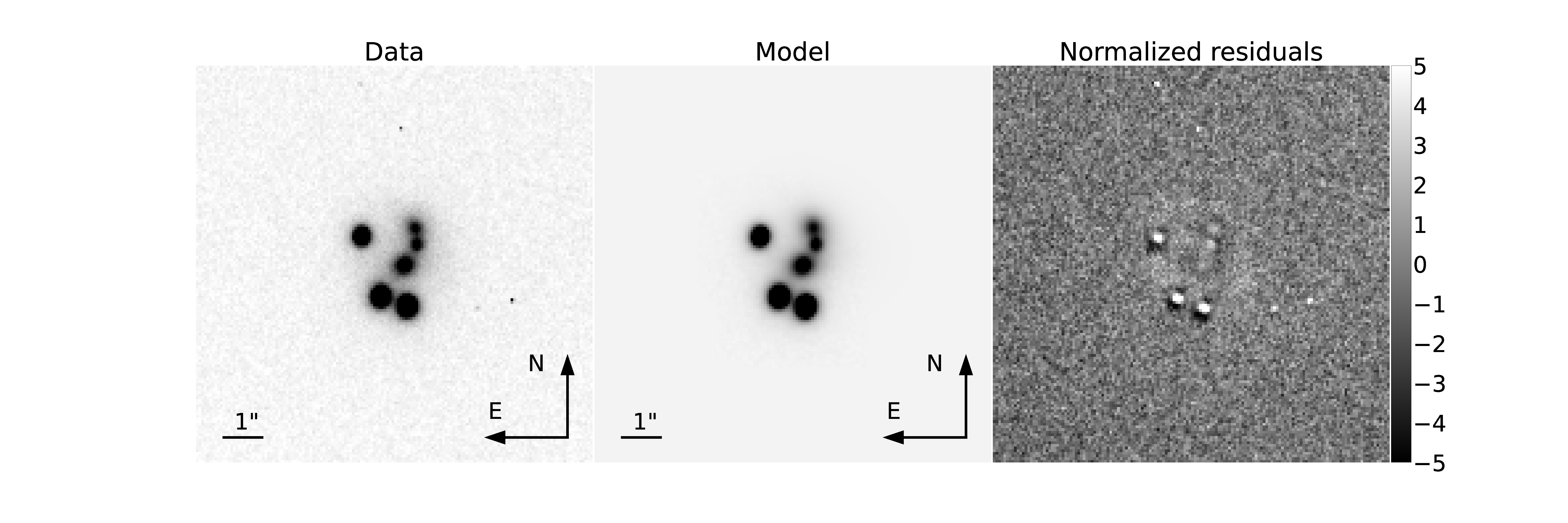}
        \caption{Observed and modeled surface-brightness distribution of HE0230$-$2130 and normalized residuals in a range of $-$5$\sigma$ to 5$\sigma$.}
        \label{fig:lightmodel}
\end{figure*}

\subsection{Sampling and analysis}
\label{sec:sampling}
We sampled the parameter space of each model class with MCMC chains, where each accepted point in the chain corresponds to one specific model realization. Each chain consists of 10,000,000 accepted points. As it is difficult to incorporate into the sampling process the fact that there is not a fifth, observable image, we use the four modeled image positions as a constraint for our models and subsequently select those model realizations that predict the correct number of images. Thus, we do not sample a distribution of candidate models, but a distribution where some models might fulfill the criterion of producing four observable images. We account not only for models that directly produce only four images, but also additional images that are too faint and are hidden in the noise of the data. 
To robustly determine those candidate models in the MCMC chains, we computed the image positions and magnifications for each model realization using the Delaunay triangulation method\footnote{In particular, we are using the Triangle package in Python: \href{https://rufat.be/triangle/}{rufat.be/triangle/}}. For this, the image plane is divided into triangles. The image plane triangles that, when mapped to the source plane, contain the mean weighted source position are iteratively refined. 

If a model produces only four quasar images, or if all additional images are below the flux limit of our data, we count it as a candidate model. We estimated the flux limit in the following way. Because of surface-brightness conservation, the minimum magnification needed for an image to be observable can be calculated via $|\mu_{\rm min}|=|\mu_{i}|\frac{K_{\rm min}}{K_{i}}$, where $\mu_{i}$ is the model-predicted magnification of image $i$, $K_{i}$ is the modeled amplitude (flux) of an observed image $i$ (that can be image A, B, C, or D), and $K_{\rm min}$ is the minimum amplitude for a light source to be above the flux limit. We conservatively chose image $i$ for which the predicted magnification limit $\mu_{\rm min}$ is the lowest. We then compared this technical magnification limit with the predicted magnification of additional images. For predicted images that are very close to the center of G2 (i.e., the innermost 3 pixels $\approx 0.2\arcsec$), we obtain a separate magnification limit, which is slightly higher due to the residuals from modeling the light of G2. The priors on the mass and shear parameters in each model are listed in Table~\ref{tab:priors}.

\section{Results and discussion}
\label{sec:results}

\subsection{Light parameters of lens and source}
We obtained astrometric positions of the QSO images from the lens and quasar light model by fitting the PSF to the observed quasar image light. Figure~\ref{fig:lightmodel} shows the results of this light fitting; here we compare the data with our model and present the normalized residuals in a range from $-$5$\sigma$ to 5$\sigma$. The residuals at the quasar positions are due to the fact that the PSF is constructed from only one star and that no corrections were performed.

Table~\ref{tab:lenslight} presents the modeled lens light parameters from fitting one S\'{e}rsic profile to both G1 and G2. Throughout the paper, we report positions in the coordinate system of CASTLES\footnote{\href{https://lweb.cfa.harvard.edu/castles/}{lweb.cfa.harvard.edu/castles} (C.S. Kochanek, E.E. Falco, C. Impey, J. Lehar, B. McLeod, H.-W. Rix)}, where we take our modeled position of image A as coordinate origin.

\begin{table}[h]
\caption{Lens light modeling results.}
\centering
                \begin{tabular}{lp{1.8cm}p{2cm}}\toprule
                        Parameter & G1 & G2 \\ \midrule \midrule \vspace{5px} 
                        $x_{\rm S}$ [\arcsec] & $-0.0801_{-0.0002}^{+0.0001}$  & $0.194_{-0.002}^{+0.002}$ \\ \vspace{5px} 
                        $y_{\rm S}$ [\arcsec] & $1.0674_{-0.0001}^{+0.0001}$  & $2.052_{-0.003}^{+0.003}$ \\ \vspace{5px} 
                        $q_{\rm S}$  & $0.64_{-0.01}^{+0.01}$ & $0.78_{-0.02}^{+0.02}$  \\ \vspace{5px} 
                        $\phi_{\rm S}$ [$^{\circ}$] & $101_{-1}^{+1}$  & $52_{-3}^{+3}$  \\ \vspace{5px} 
                        $A_{\rm S}$ & $334_{-1}^{+1}$  & $199_{-1}^{+1}$  \\ \vspace{5px} 
                        $r_{\rm eff}$ [\arcsec] & $0.422_{-0.004}^{+0.005}$  & $0.424_{-0.006}^{+0.006}$ \\ \vspace{5px} 
                        $n_{\rm S}$ & $2.99_{-0.06}^{+0.07}$  & $2.74_{-0.07}^{+0.08}$ \\
             \toprule  
                \end{tabular}
            
                \ \
        \caption*{\textbf{Notes: } G1 and G2 are each modeled with one S\'{e}rsic profile. The position angle is measured east of north. The modeled position of image A is set as the origin of the coordinate system.}
        \label{tab:lenslight}
\end{table}

G1 and G2 show regular isophotes as both can be fit well with one S\'{e}rsic profile each. Given this and the fact that we need a relative velocity of between G1 and G2 of around 1000 km/s (see Sect.~\ref{sec:intro}), strong dynamical interaction between the two lensing galaxies is not likely, although we cannot rule it out.

Table~\ref{tab:astrometry} presents our modeled image positions and compares them with those measured by Gaia \citep{Brown2018} and by CASTLES with HST data. In this table, we also report measured fluxes from our models. As a magnitude zero point is missing for the Magellan data, we calibrated one using the magnitudes of stars in the field measured in the Legacy Survey DR10 and also calculated their fluxes in the AB magnitude system. The fluxes from CASTLES are reported in the Vega system.

\begin{table*}[h]
\caption{Astrometric positions and fluxes of quasar images.}
\centering
\fontsize{7}{8}\selectfont
        \begin{tabular}{llccc}\toprule
         image & &  \GLEE & CASTLES & Gaia \\\midrule \vspace{5px} 
                 & x [\arcsec] & $\equiv$ 0 & $\equiv$ 0 & $\equiv$ 0 \\ \vspace{5px} 
                A & y [\arcsec] & $\equiv$ 0 & $\equiv$ 0 & $\equiv$ 0 \\
                 & flux & 19.34 (AB) & 19.02 (Vega,F814W) & $-$ \\
                & & 18.97 (Vega,$i$) & & \\\midrule \vspace{5px} 
                 & x [\arcsec] & $-0.695^{+0.002}_{-0.002}$ & $-0.698^{+0.003}_{-0.003}$ & $-$0.697 \\ \vspace{5px} 
                B & y [\arcsec] & $0.258^{+0.002}_{-0.002}$  & $0.256^{+0.003}_{-0.003}$ & 0.258 \\ 
                 & flux & 19.47 (AB) & 19.22 (Vega,F814W) &$-$ \\
                & & 19.10 (Vega,$i$) & & \\\midrule \vspace{5px} 
                & x [\arcsec] & $-1.192^{+0.002}_{-0.002}$ & $-1.198^{+0.005}_{-0.005}$ & $-$1.198 \\ \vspace{5px} 
                C & y [\arcsec] & $1.827^{+0.002}_{-0.002}$& $1.828^{+0.003}_{-0.003}$  & 1.832 \\
                 & flux & 19.91 (AB) & 19.59 (Vega,F814W) &$-$ \\
                & & 19.54 (Vega,$i$) & & \\\midrule \vspace{5px} 
                & x [\arcsec] & $0.261^{+0.003}_{-0.003}$ & $0.244^{+0.007}_{-0.007}$  & $-$\\ \vspace{5px} 
                D & y [\arcsec] & $1.610^{+0.003}_{-0.003}$ & $1.624^{+0.007}_{-0.007}$ &$-$ \\
                 & flux & 21.59 (AB) &21.21 (Vega,F814W) &$-$ \\
                 & & 21.22 (Vega,$i$) & &  \\\midrule
        \end{tabular}

                \vspace{0.2cm}
        \caption*{\textbf{Notes: } Positional values and uncertainties are given in arcseconds and are shifted into the CASTLES reference frame. The median modeled position of image A is set as the origin of the coordinate system. There is no measured position of image D from Gaia as it is below the detection limit of the Gaia satellite. The fluxes in the Magellan $i$-band are reported in the AB magnitude system and converted to Vega magnitudes (noted as Vega,i) using Table~1 in \citet{Blanton2007}. The CASTLES fluxes are those of the HST F814W band and reported in the Vega system (noted as Vega,F814W). The transmission curves of Magellan and the F814W filter from HST WFPC2 have different shapes, which needs to be taken into account when comparing the GLEE with the CASTLES magnitudes.}

        \label{tab:astrometry}
\end{table*}

We find that our modeled image positions agree within 6 mas in the $x$-direction and 5 mas in the $y$-direction with Castles and Gaia. The root-mean-square (rms) offset with Gaia is $\sim$2 mas in both $x$- and $y$-directions. We therefore added 2\,mas in quadrature to the statistical uncertainties from the MCMC chain to obtain the total uncertainty in the astrometry of the quasars. Given this good agreement, we did not conduct corrections to the PSF to reduce the residuals at the quasar images seen in Fig.~ \ref{fig:lightmodel}. We exclude only image D in this comparison, which shows a greater offset to the position for Castles because it is the faintest image and is partially blended with the light of G2. 

\subsection{Candidate models}

We present the final mass and shear parameters of all model classes in Table~\ref{tab:results_param}. We list the distributions of parameters for both the full chain and the candidate model realizations. Table~\ref{tab:results} shows an overview of our results. The modeled position of image A is set as the origin of the coordinate system.

\begin{table}[h]
\centering
\caption{Overview of the candidate fraction of all model classes.}
                \begin{tabular}{l|p{1.8cm}|p{1.5cm}}\toprule
                        Model & correct \# of images & $f_{\rm cand}$ \\ \midrule \midrule
                        \phantom{c }PL1 + \phantom{c}PL2 & $\times$  &  0\% \\ \rule{0pt}{2ex}
                        \phantom{c}PL1 + \phantom{c}PL2 + $\gamma_{\rm ext}$ & $\times$  & 0\% \\ \rule{0pt}{2ex}
                        \phantom{c}PL1 + cPL2 & $\times$  &  0\%  \\ \rule{0pt}{2ex}
                        \phantom{c}PL1 + cPL2 + $\gamma_{\rm ext}$ & $\checkmark$  & 0.26\% \\ \rule{0pt}{2ex}
                        cPL1 + cPL2 & $\times$  &  0\%  \\ \rule{0pt}{2ex}
                        cPL1 + cPL2 + $\gamma_{\rm ext}$ & $\checkmark$  & 0.26\% \\ \rule{0pt}{2ex}
                        \phantom{c}PL1 + \phantom{c}PL2 + \phantom{r}SIS & $\times$  & 0\% \\ \rule{0pt}{2ex}
                        \phantom{c}PL1 + \phantom{c}PL2 + \phantom{r}NIS & $\times$ & 0\% \\ \rule{0pt}{2ex}
                        \phantom{c}PL1 + \phantom{c}PL2 + rSIS & $\checkmark$  & 0.03\%  \\ \rule{0pt}{2ex}
                        \phantom{c}PL1 + \phantom{c}PL2 + rNIS & $\checkmark$  &  1.7\%  \\ \rule{0pt}{2ex}
                        \phantom{c}PL1 + \phantom{c}PL2 + rSIS + $\gamma_{\rm ext}$ & $\checkmark$  & 0.4\%  \\ \rule{0pt}{2ex}
                        \phantom{c}PL1 + \phantom{c}PL2 + rNIS + $\gamma_{\rm ext}$ & $\checkmark$  & 12\% \\ 
              \bottomrule          

                \end{tabular}
                                    \vspace{0.2cm}
        \caption*{\textbf{Notes: } Candidate models are model realizations that predict four instead of five quasar images, or where additional images are below the flux limit of the data. $f_{\rm cand}$ is the weighted percentage of candidate model realizations in the MCMC chain. In this computation, we take the statistical weights of the MCMC chain into account.}
        \label{tab:results}
\end{table}

Of the 12 model classes we considered, 6 can produce models that match the observed image multiplicity. We find that the simplest models, where G1 and G2 are parametrized by a singular PL profile with or without external shear, consistently produce exactly one additional image, as expected from this lensing configuration. These additional images are located about 0.5\arcsec west of G2, with a magnification that should make it observable. Figure~\ref{fig:2pl_5thimg} shows the 1$\sigma$ and 2$\sigma$ contours of the predicted positions of image E in the PL1 + PL2, the PL1+cPL2+$\gamma_{ext}$, and the PL1+PL2+rNIS model classes.

In the class PL1 + cPL2 + $\gamma_{\rm ext}$, 0.26\% of the models  produce the correct, observed image multiplicity, while for PL1 + cPL2 (same parameterization for G1 and G2 but without external shear), there are no models that match the observation. Similarly, for the model class cPL1 + cPL2 + $\gamma_{\rm ext}$, 0.26\% of the models fit the observations, while there are no candidate models in cPL1 + cPL2. 

A similar effect might be produced by placing an SIS or NIS mass distribution close to the saddle point of image E on the lens plane. This mass can be interpreted as DM substructure or an unobservable dark galaxy. We find that by simply placing a SIS or NIS clump without positional prior, they tend to move away from the region close to the saddle point and predominantly end up west of image D. Therefore, these two model classes are unable to suppress image E and we introduce the rSIS and rNIS mass clumps, which have a Gaussian prior on their centroids to keep them closer to the region west of G2.

G1 and G2 are parametrized as singular PL profiles. Of the PL1 + PL2 + rSIS models, 0.03\% do not predict a fifth image. This is also the case for 1.7\% of the PL1 + PL2 + rNIS models.

In our analysis, we assumed the lensing galaxies to be close to isothermal based on the analysis of SLACS lenses (see Sect.~\ref{sec:intro} with a Gaussian prior centered on $\gamma=2.08)$. We tested the influence of this prior on our results by using a shallower slope of $\gamma=1.8$. We find that the exact candidate fraction is sensitive to the prior on $\gamma$ but our conclusions remain unchanged: a singular PL model for G1 and G2 cannot explain the observed image multiplicity, while models with cored PLs or a rSIS and rNIS model fit the observations.

Finally, we analyzed the PL1 + PL2 + rSIS and  PL1 + PL2 + rNIS models with the presence of external shear. We find that the addition of external shear significantly increases the fraction of candidates to 0.4\% in the  PL1 + PL2 + rSIS+$\gamma_{\rm ext}$ case, and to 12\% for the  PL1 + PL2 + rNIS+$\gamma_{\rm ext}$ class.
In the following, we provide a detailed description and interpretation of the results for each model class that produces the observed four images.

\subsubsection{PL1 + cPL2 + $\gamma_{\rm ext}$}
\begin{figure*}[h]
\centering
        \includegraphics[scale=1.2]{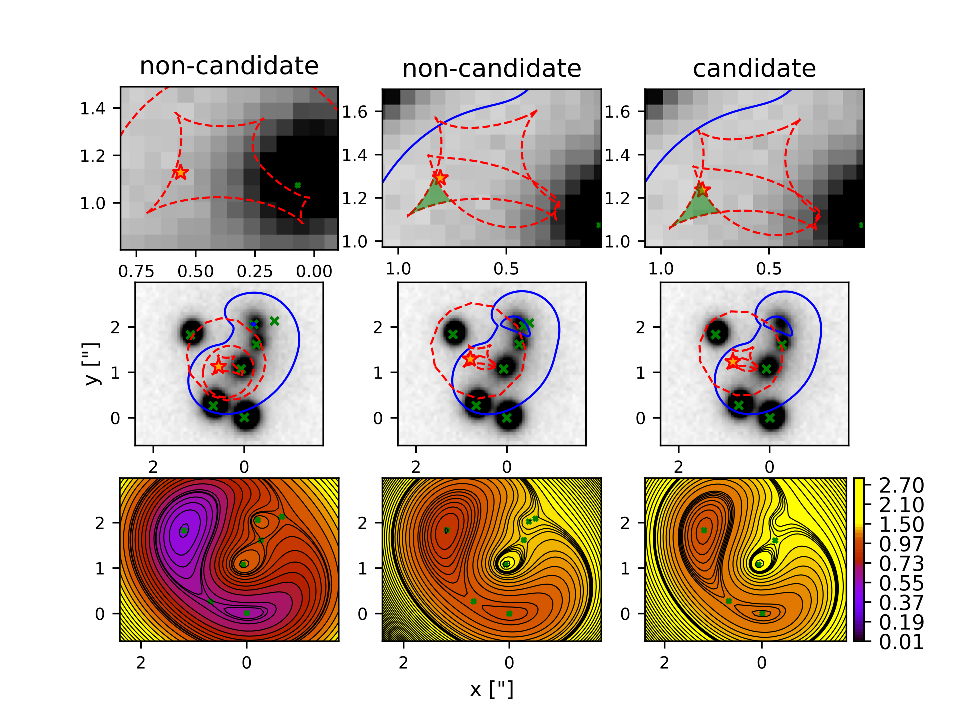}
        \caption{Change of critical curves and caustics between noncandidate models (left and middle columns) and one candidate model (right column). The top and middle rows show the critical curves (blue, solid lines) and caustic curves (red, dashed lines). The bottom row shows the Fermat potential and contours with the color scale in units of ${\textrm{arcsec}}^2$.
        As the core radius of G2 increases and the source position crosses the radial caustic into the shaded green region, the time-delay saddle point associated with image E merges with the maximum of G2 and image E disappears. 
        The source position is plotted as a yellow star.  The
predicted image positions are shown as green crosses in each panel.}
        \label{fig:g2core_srcpos}
\end{figure*}

In this model class, G1 is described by a singular PL profile, while G2 is parametrized as a cored PL profile. Galaxies that are part of clusters often exhibit a core, while  a core is not common for isolated field galaxies. An additional external shear component is necessary to reproduce the observed image multiplicity, as the PL1 + cPL2 model class always predicts an additional image.
By adding a core to the mass distribution of G2, the saddle point of image E merges with the maximum of G2. In Fig.~\ref{fig:g2core_srcpos}, we plot the Fermat potential of two noncandidate and one candidate model; that is, before and after merging of the saddle with the maximum. The difference between candidate and noncandidate is also evident in the source position with respect to the caustic curves in the source plane: as the radial caustic shrinks and changes in shape, the source crosses the radial caustic into the shaded green region, and the image multiplicity of this model changes.\footnote{The predicted image near the center of G1 is highly demagnified given its singular mass profile at the center.  For stability in numerical computations, we cannot have a singularity in the mass profile, and have therefore a softening radius of $(10^{-4})\arcsec$, which generates a central image that is highly demagnified.} Furthermore, the Fermat potential is flattened in the vicinity of the vanished image.

Figure~\ref{fig:g2corner} shows the posterior distribution of selected lens-mass and shear parameters. The distributions of candidate models are plotted in red contours, and the distribution of the whole chain (i.e., both candidate and noncandidate models) is plotted in black contours. We find that candidate models have a very round mass distribution of G2 with $q_{\rm G2,cand}=0.97_{-0.05}^{+0.02}$, while the general distribution of the chain is more uniformly distributed over the whole prior range: $q_{\rm G2,all}=0.77_{-0.12}^{+0.15}$. Similarly, the general distribution of $r_{\rm c, G2}$ is skewed towards a vanishing core radius (i.e., a singular PL profile). In order to produce only four images, a core radius of $r_{\rm c, G2, cand}=(0.14_{-0.04}^{+0.04})\arcsec$ is needed. This corresponds to a physical core size of $r_{\rm c, G2, cand}=(0.88_{-0.25}^{+0.25})$ kpc, which is in good agreement with typical observed core sizes in elliptical galaxies (see Sect.~\ref{sec:intro}). The candidate models are all found in the tail of the distribution. The shear strength of candidate models is $\gamma_{\rm ext,cand}=0.08_{-0.02}^{+0.02}$. This is a typical value for quadruply lensed quasars as shown by \citealt{Luhtaru2021}.

\subsubsection{cPL1 + cPL2 + $\gamma_{\rm ext}$}
Now we parametrize both G1 and G2 with a cored PL profile. As for the PL1 + cPL2 + $\gamma_{\rm ext}$ case, we need external shear to reproduce the correct image multiplicity. The mechanism by which image E is eliminated is the same as in the PL1 + cPL2 + $\gamma_{\rm ext}$ model class: the saddle point of image E merges with the maximum of G2.
We show the posterior distributions of all models compared to the candidates in Fig~\ref{fig:g1g2corner}. Again, a clear tendency of the candidate models is a high $q_{\rm G2,cand}$ compared to the more uniform, general distribution. The distribution of candidate core radii are $r_{\rm c, G1, cand}=(0.16_{-0.06}^{+0.04})$\arcsec $\sim (1.00_{-0.38}^{+0.25})$ kpc and $r_{\rm c, G2, cand}=(0.13_{-0.04}^{+0.04})$\arcsec $\sim (0.82_{-0.25}^{+0.25})$ kpc, respectively. The external shear has a distribution of $\gamma_{\rm ext,cand}=0.08_{-0.02}^{+0.01}$.

\subsubsection{PL1 + PL2 + rSIS and PL1 + PL2 + rNIS}
The physical motivation behind the model classes with an SIS or NIS profile is to simulate a dark mass distribution in the plane of the lensing galaxies; for instance, DM substructure or a dwarf galaxy. The PL1 + PL2 + SIS and PL1 + PL2 + NIS model classes cannot reproduce the correct number of images as the centroids of the mass clumps are moving away from the critical region where the saddle of image E can be merged with the maximum of G2. We assume a Gaussian prior on the centroid of the SIS/NIS clump to keep it close to the critical region shown in Fig~\ref{fig:2pl_5thimg}. As mentioned earlier, we refer to these as restricted SIS and NIS profiles, or rSIS and rNIS. 
While the PL1 + PL2 + SIS and PL1 + PL2 + NIS model classes do not predict models with only four images, 0.03\% of PL1 + PL2 + rSIS models produce no fifth image. For the PL1 + PL2 + rNIS model class, this percentage is 1.7\%. We can see in Fig~\ref{fig:rsiscorner} and Fig~\ref{fig:rniscorner}, that, with the presence of an additional mass component close to G2, the Einstein radius $\theta_{\rm E,G2}$ decreases. The PL1 + PL2 + rSIS candidate models predict an especially low Einstein radius of G2 with $\theta_{\rm E,G2,cand}=(0.19_{-0.03}^{+0.02})\arcsec$, while it is higher for the candidates in the rNIS model: $\theta_{\rm E,G2,cand}=(0.33_{-0.14}^{+0.11})\arcsec$. The Einstein radius of the rSIS in candidate models is $\theta_{\rm E,rSIS,cand}=(0.18_{-0.05}^{+0.05})\arcsec$, which is of a similar size to $\theta_{\rm E,G2,cand}$. The PL1 + PL2 + rNIS candidates have an rNIS Einstein radius of $\theta_{\rm E,rNIS,cand}=(0.13_{-0.09}^{+0.12})\arcsec$, which is smaller than $\theta_{\rm E,G2,cand}$ in this model class. It is also smaller than the inferred core radius of $r_{\rm c,rNIS,cand}=(0.34_{-0.13}^{+0.19})\arcsec$.

While subhalos are modeled with a Navarro-Frenk-White (NFW; \citealt{Navarro1996,Navarro1997}) profile  in many studies, we chose to model the subhalo with an isothermal sphere profile to keep the number of free parameters as small as possible. Other studies use a Pseude-Jaffe (PJ; \citealt{Dalal2001,Vegetti2014b,Despali2017}) profile, which is an isothermal profile with a truncation radius. We conducted a quick model run and find that both the NFW and PJ models are also able to suppress image E with a similar candidate fraction in the MCMC chains. The success of a model with an additional mass clump is therefore independent of the specific choice of mass profile. Furthermore, we do not have any observational constraints on the position of the subhalo along the line of sight, which was shown to be degenerate with the mass of the subhalo by \citealt{Vegetti2014} and \citealt{Wagner2017}. The lensing effect more robustly depends on the projected mass of the subhalo. \citet{Minor2017} showed that the projected mass can be robustly inferred even if the mass density profile of the subhalo is not known. We discuss the projected masses of the rNIS clumps in Sect.~\ref{sec:subhalo_mass}.

\subsubsection{PL1 + PL2 + rSIS + $\gamma_{\rm ext}$ and PL1 + PL2 + rNIS + $\gamma_{\rm ext}$}
\begin{figure}
\centering
        \includegraphics[scale=0.6]{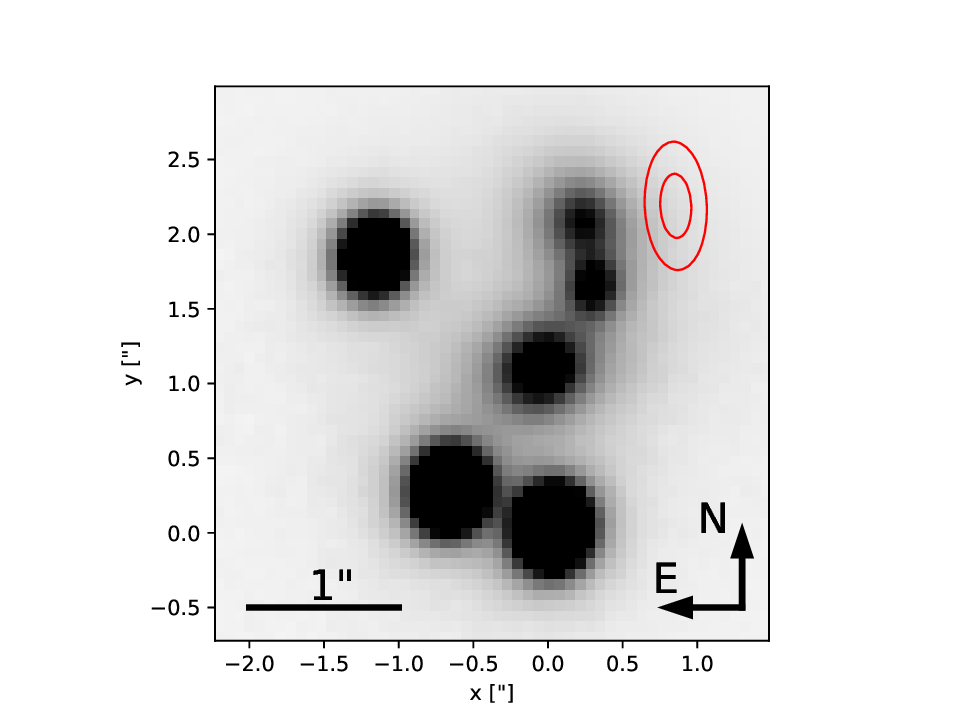}
        \caption{1$\sigma$ and 2$\sigma$ contours of the mass centroid of the candidate rNIS clumps in the PL1 + PL2 + rNIS model class. These candidate models predict only four quasar images, or additional images that are too faint to be observed (i.e., below the flux limit of the data).}
        \label{fig:rnis+s_pos}
\end{figure}

By adding external shear to the PL1 + PL2 + rSIS and PL1 + PL2 + rNIS model classes, the fraction of candidate models increases substantially. The effect of adding external shear to our models can be seen in the corner plots in Fig.~\ref{fig:rniscorner} and \ref{fig:rnis+scorner}. The physical meaning of external shear has recently been challenged, as shear values inferred from strong lensing are inconsistent with independent measurements from, for example, weak lensing \citep{Etherington2023}. The fact that the inclusion of external shear in our models generally increases the candidate fraction may therefore mean that there is some complexity in this lensing system that cannot be accounted for with just two (cored) PL profiles for G1 and G2. This is not surprising as G1 and G2 are possibly embedded in a common DM halo. In particular, 12\% of the models of the PL1 + PL2 + rNIS + $\gamma_{\rm ext}$ class produce the correct number of images. The 1$\sigma$ and 2$\sigma$ contours of the PL1 + PL2 + rNIS centroids are plotted on the modeled image in Fig.~\ref{fig:rnis+s_pos}. This shows that the mass clumps in models with the correct image multiplicity are located about 0.5\arcsec northwest of G2. The posterior distributions in Fig~\ref{fig:rnis+scorner} show that the PL1 + PL2 + rNIS + $\gamma_{\rm ext}$ candidates have an rNIS Einstein radius of $\theta_{\rm E,rNIS+\gamma_{\rm ext},cand}=(0.24_{-0.16}^{+0.17})\arcsec$ and a core radius of $r_{\rm c,rNIS+\gamma_{\rm ext},cand}=(0.24_{-0.12}^{+0.16})\arcsec$. The external shear strength is $\gamma_{\rm ext,cand}=0.05_{-0.01}^{+0.02}$. As in the PL1 + PL2 + rSIS model, the Einstein radius $\theta_{\rm E,rNIS+\gamma_{\rm ext},cand}$ is of similar size to $\theta_{\rm E,G2,cand}$, which implies that the mass clump should be sufficiently massive to be luminous.

\subsection{Model comparison}
As there are multiple model classes that are able to produce the correct number of observed images, we investigated which of the model classes is more likely by comparing their likelihoods. Rather than using the minimum $\chi^2_{\rm im}$ for the comparison, we calculated the Bayesian information criterion (BIC) to weigh the model according to both their complexity and goodness of fit. The BIC is defined as
\begin{equation}
    \textrm{BIC}=k\ln(N_{\rm data}) - 2\ln(\hat{\mathcal{L}}),
\end{equation}
where $k$ is the number of free parameters, $N_{\rm data}$ is the number of data points, and $\hat{\mathcal{L}}$ is the maximum likelihood of the candidate models. The relative probability of a model M with BIC compared to the most likely model $\textrm{M}^{*}$ with $\textrm{BIC}^{*}$ is
\begin{equation}
    p(\textrm{M})/p(\textrm{M}^{*})=\textrm{exp}(-\frac{\textrm{BIC}-\textrm{BIC}^{*}}{2})\times \frac{f_{\rm cand,M}}{f_{\rm cand,{M}^{*}}},
    \label{eq:relprob}
\end{equation}
where $f_{\rm cand}$ is the weighted fraction of candidate models in the MCMC chain. In terms of free parameters $k,$ we have the free lens galaxy model parameters, plus two for the source position. Because we calculate the flux limit that is specific for each model class and which we use to select candidate models, we count the source intensity as an additional free parameter and the image amplitude $K_{i}$ (see Sect.\ref{sec:sampling}) as an additional data point to the eight data points from the four image positions. Additionally, we count the Gaussian priors as both free parameter and data point. Table~\ref{tab:BIC} lists the BIC values and relative probabilities of our six candidate models.

\begin{table*}[h]
\caption{Bayesian comparison of candidate model classes.}
\centering
        \begin{tabular}{llllccc}\toprule
                model & $k$ & $N_{\rm data}$ & $f_{\rm cand}$ & min. $\chi^2_{\rm im}$ & BIC & $p(\textrm{M})/p(\textrm{M}^{*})$ \\\midrule\midrule \vspace{5px} 
                \phantom{c}PL1 + cPL2 + $\gamma_{\rm ext}$ & 14 & 13 & 0.26\% & 1.32 & 37.23 & 0.94  \\ \vspace{5px} 
                cPL1 + cPL2 + $\gamma_{\rm ext}$ & 15 & 13 & 0.26\% & 0.54 & 39.01 & 0.38   \\ \vspace{5px} 
                \phantom{c}PL1 + \phantom{c}PL2 + rSIS & 14 & 15 & 0.03\% & 1.60 & 39.51 & 0.04  \\ \vspace{5px} 
                \phantom{c}PL1 + \phantom{c}PL2 + rNIS & 15 & 15 & 1.7\% & 0.12 & 40.80 & 1 \\  \vspace{5px}
                \phantom{c}PL1 + \phantom{c}PL2 + rSIS + $\gamma_{\rm ext}$ & 16 & 15 & 0.4\%  & 0.13 & 43.46 & 0.07 \\ \vspace{5px}
                \phantom{c}PL1 + \phantom{c}PL2 + rNIS + $\gamma_{\rm ext}$ & 17 & 15 & 12\%  & 0.03  & 46.07 & 0.53 \\ \bottomrule
        \end{tabular}

                \vspace{0.2cm}
        \caption*{\textbf{Notes: } $k$ is the number of free parameters, $N_{\rm data}$ is the number of data points, and $f_{\rm cand}$ is the weighted fraction of candidate models in the MCMC chain. The relative probability $p(\textrm{M})/p(\textrm{M}^{*})$ is calculated via Eq.~\ref{eq:relprob} and is then renormalized.}
        \label{tab:BIC}
\end{table*}

The most likely model class is PL1 + PL2+ rNIS, which is slightly favored over the model class PL1 + cPL2 + $\gamma_{\rm ext}$, which has a relative probability of 94\%. Although cPL1 + cPL2 + $\gamma_{\rm ext}$ has a lower $\chi^2_{\rm im}$ and a similar fraction of candidates compared to PL1 + cPL2 + $\gamma_{\rm ext}$, the latter is simpler in terms of model complexity (and has fewer free parameters) and is therefore favored by the BIC. In addition to this model ranking based on Bayesian statistics, we checked the physical properties, such as the mass-to-light ratios of the lens galaxies and the mass of the mass clumps in the PL1 + PL2+ rNIS model class in the following two subsections in order to show which models are physically reasonable.

\subsection{Plausibility of inferred parameter values}
To check the plausibility of the model parameter values of our two favored model classes, we estimate the mass-to-light ratio $\frac{M}{L}$ of G1 and G2 for both classes. We obtain the luminosity $L$ by numerically integrating the modeled S\'{e}rsic light profile (see Eq.~\ref{eq:sersic}) within the effective radius $r_{\rm eff}$ using a circular aperture. From the integrated fluxes, we obtain apparent magnitudes of $m_{{\rm G1,AB},i}=20.39$ and $m_{{\rm G2,AB},i}=20.79$ in the observed $i$-band. The two lensing galaxies are located at redshifts of 0.523 and 0.526, respectively, which approximately corresponds to rest-frame $g$-band. We conducted a K correction \citep{Hogg2002} to transform the apparent magnitudes in the $i$-band to apparent magnitudes in the rest-frame $g$-band via $m_{{\rm AB},g} = m_{{\rm AB},i} - K_{gi}$ with $K_{gi}=-0.64$. We used the galaxy spectra from \citealt{Eigenbrod2006} and fitted an ETG spectral template. The rest-frame apparent magnitudes are therefore $m_{\rm G1,AB}=21.03$ and $m_{\rm G2,AB}=21.43$, which convert to $L_{\rm G1}=3.83\times10^{10} L_{\odot}$ and $L_{\rm G2}=2.69\times10^{10} L_{\odot}$. We infer the mass by integrating the $\kappa_{\rm PL}$ profile of the respective model class within the same effective radius as the light. From a distribution of candidate models, we find $M_{\rm G1,cPL2}=8.1^{+0.9}_{-0.7}\times10^{10}  M_{\odot}$ and $M_{\rm G2,cPL2}=3.5^{+0.6}_{-0.5}\times10^{10}  M_{\odot}$ for the PL1+cPL2+$\gamma_{\rm ext}$ model class, and $M_{\rm G1,rNIS}=7.5^{+0.6}_{-0.5}\times10^{10}  M_{\odot}$ and $M_{\rm G2,rNIS}=2.5^{+2.0}_{-0.7}\times10^{10}  M_{\odot}$ for the PL1+PL2+rNIS model class. For G1, we therefore find mass-to-light ratios of $\Big(\frac{M}{L}\Big)_{\rm G1,cPL2}=2.1^{+0.2}_{-0.2}$ and $\Big(\frac{M}{L}\Big)_{\rm G1,rNIS}=2.0^{+0.2}_{-0.1}$. For G2, we find $\Big(\frac{M}{L}\Big)_{\rm G2,cPL2}=1.3^{+0.2}_{-0.2}$ and $\Big(\frac{M}{L}\Big)_{\rm G2,rNIS}=0.94^{+0.7}_{-0.2}$. Galaxies with a similar magnitude are expected to have $(\frac{M}{L})_{g} \sim 1-5$ \citep{Kauffmann2003, Bell2003}, and therefore our inferred masses are plausible. We note that the mass of G2 is degenerate with the mass of the NIS subhalo, which means that $M_{\rm G2,rNIS}$, and therefore the mass-to-light ratio, might be underestimated.

\subsection{Detection of DM substructure?}
\label{sec:subhalo_mass}
The model class PL1+PL2+rNIS, which is preferred by the BIC, implies the presence of a mass clump about 0.5\arcsec\ to 1\arcsec\ northwest of G2. We subsequently investigated the nature of the predicted rNIS clumps in the PL1 + PL2 + rNIS model class by comparing their mass with the expected mass range for subhalos from simulations. For a total mass estimate of the rNIS clump, we calculated its effective subhalo lensing mass \citep{Minor2017}, which is robust to changes in the mass density profile and is a good estimate of the total mass of the subhalo. The effective subhalo lensing mass is the mass enclosed within a subhalo perturbation radius, scaled by the slope of the host galaxy, which we assume to be G2.

\begin{figure}
\centering
        \includegraphics[scale=0.6]{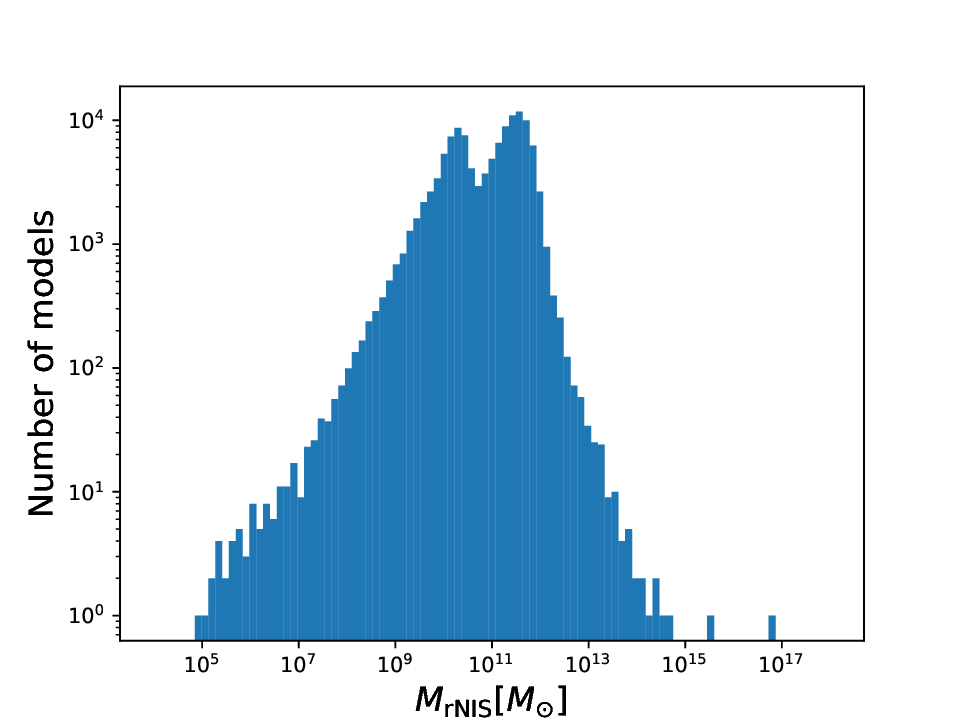}
        \caption{Logarithmic distribution of the masses of rNIS clumps (in solar masses) in the candidate models of the PL1 + PL2 + rNIS + $\gamma_{\rm ext}$ model class.}
        \label{fig:nis_mass}
\end{figure}

We show the distribution of $M_{\rm rNIS}$ of all the candidate models in the chain on a logarithmic scale in Fig.~\ref{fig:nis_mass}. The distribution spans a mass range of $\sim8\times10^4 - 6\times10^{16} M_{\odot}$. Approximately 2\% of all candidate models predict $10^6 M_{\odot} \leq M_{\rm rNIS}\leq 10^9 M_{\odot}$, which is the range of predicted masses for DM substructure or dark dwarf-galaxies in simulations \citep{Dalal2001,Simon2007,Diemand2008,Vegetti2012}. Furthermore, $\sim$14\% of all candidate models predict $M_{\rm rNIS}\leq 10^{10} M_{\odot}$. For halos with mass above $10^9 M_{\odot}$, we would expect them to host a galaxy and therefore be luminous. The large fraction of rNIS models with $M_{\rm rNIS}> 10^{10}$ results from the combination of a small Einstein radius of G2 and a large distance of the subhalo to G2. This leads to a bigger subhalo perturbation radius and therefore to a higher effective subhalo lensing mass. We conclude that the presence of unobserved dark matter structure close to the lens is unlikely to be responsible for the missing image because the effective subhalo masses are mostly too large compared to the expected masses for subhalos from simulations. It is possible that the rNIS tries to fit to the group halo of the galaxy group that G1 and G2 are part of, but the lack of high-resolution imaging data prevents us from further investigating this point at present.
Our preferred explanation is therefore the presence of a core in G2, which changes the landscape of critical curves and caustics, and ultimately eliminates the fifth quasar image.

\section{Summary}
\label{sec:summary}
We modeled and analyzed the strongly lensed quasar HE0230$-$2130 using ground-based imaging data from the Magellan telescope. This lensing system shows a peculiar configuration with two lensing galaxies at similar redshift and four quasar images. The lensing galaxies are possibly part of a larger galaxy group and it is possible that the two lensing galaxies are dynamically associated or are embedded in an overall DM halo. As we expect five quasar images for such a lensing configuration, our aim is to find models that produce only four images. After we obtained the image positions by modeling the lens and quasar light in the observed image, we use them to constrain the mass parameters of our models. The straightforward approach of modeling both lens galaxies with a singular PL profile agrees with our expectations and always results in one additional predicted image, which is not observable in the data. We test a dozen different model parametrizations where we add core radii to the PL profiles describing G1 and G2, and/or we add a singular or cored isothermal sphere profile at the same redshift as G1 and G2, mimicking a dark mass distribution. We find that introducing a core to G2, or indeed to G1 and G2, can explain the missing image, as can a cored or singular isothermal sphere close to the predicted position of the missing image. In all of these cases, the saddle of the missing image merges with the maximum of G2, such that only four observable quasar images remain. All model classes consistently predict an external shear strength between 0.06 and 0.08. Adding external shear to the models generally increases the candidate fraction. This could mean that this system is potentially not dynamically relaxed, as the external shear parameter has  recently been shown  to compensate for missing model complexity.
We compared the likelihoods of each candidate model class with the BIC, taking into account the fraction of candidate models. The most likely model class is cPL1 + cPL2 + rNIS, where 1.7\% of all models produce the correct number of images, but the median mass of these NIS clumps is $\sim2.9\times10^{10} M_{\odot}$, which is one order of magnitude larger than the upper mass limit of the predicted subhalos from simulations. Only about 2\%\ of the candidate models have clumps in the predicted mass range of $10^6 M_{\odot} \leq M_{\rm rNIS}\leq 10^9 M_{\odot}$ from simulations. Most of the clumps are too massive to be interpreted as undetected dark matter substructure. We therefore favor the explanation that a core in the mass distribution of G2 is responsible for the missing image.

We conclude that we find models that can describe this peculiar lensing system. Our favored explanation is the presence of a core in G2 in combination with external shear. Follow-up high-resolution and high-signal-to-noise-ratio imaging observations, which could display the quasar host galaxy as arcs or give us a better view of the region around G2, would help to further constrain our models and shed light on the unusual nature of HE0230$-$2130.

\begin{acknowledgements}
We thank R.~Levinson for reduction of the imaging data, and S.~Burles for the color image. We are grateful to the referee C.~D.~Fassnacht for constructive and detailed comments that improved the presentation and clarity of our work. SE and SHS thank the Max Planck Society for support through the Max Planck Research Group and the Max Planck Fellowship for SHS. This work is supported in part by the Deutsche Forschungsgemeinschaft (DFG, German Research Foundation) under Germany's Excellence Strategy -- EXC-2094 -- 390783311. SS acknowledges financial support through grants PRIN-MIUR 2017WSCC32 and 2020SKSTHZ.

This work made use of \texttt{NumPy} \citep{Oliphant2015}, \texttt{SciPy} \citep{Jones2001}, \texttt{astropy} \citep{AstropyCollaboration2018}, and \texttt{matplotlib} \citep{Hunter2007}.

\end{acknowledgements}

\bibliographystyle{aa}
\bibliography{bibliography}

\begin{thebibliography}{97}
\expandafter\ifx\csname natexlab\endcsname\relax\def\natexlab#1{#1}\fi

\bibitem[{Agnello {et~al.}(2017)Agnello, Lin, Buckley-Geer, Treu, Bonvin,
  Courbin, Lemon, Morishita, Amara, Auger, Birrer, Chan, Collett, More,
  Fassnacht, Frieman, Marshall, McMahon, Meylan, Suyu, Finley, Kochanek,
  Makler, Martini, Morgan, Ostrovski, Schechter, Tucker, Wechsler, Abbott,
  Abdalla, Allam, Benoit-Lévy, Bertin, Brooks, Burke, Rosell, Kind, Carretero,
  Crocce, Cunha, da~Costa, Desai, Dietrich, Eifler, Flaugher, Fosalba,
  García-Bellido, Gaztanaga, Goldstein, Gruen, Gruendl, Gschwend, Gutierrez,
  Honscheid, James, Kuehn, Kuropatkin, Li, Plazas, Romer, Sanchez, Schindler,
  Schubnell, Sevilla-Noarbe, Smith, Smith, Sobreira, Suchyta, Swanson, Tarle,
  Thomas, \& Walker}]{Agnello2017}
Agnello, A., Lin, H., Buckley-Geer, L., {et~al.} 2017, Mon. Not. R. Astron.
  Soc, 000, 29

\bibitem[{Anguita {et~al.}(2008)Anguita, Faure, Yonehara, Wambsganss, Kneib,
  Covone, \& Alloin}]{Anguita2008}
Anguita, T., Faure, C., Yonehara, A., {et~al.} 2008, A\&A, 481, 615

\bibitem[{Auger {et~al.}(2010)Auger, Treu, Bolton, Gavazzi, Koopmans, Marshall,
  Moustakas, \& Burles}]{Auger2010}
Auger, M.~W., Treu, T., Bolton, A.~S., {et~al.} 2010, Astrophysical Journal,
  724, 511

\bibitem[{Barkana(1998)}]{Barkana.1998}
Barkana, R. 1998, The Astrophysical Journal, 502, 531

\bibitem[{Barnabe {et~al.}(2011)Barnabe, Czoske, Koopmans, Treu, \&
  Bolton}]{Barnabe2011}
Barnabe, M., Czoske, O., Koopmans, L. V.~E., Treu, T., \& Bolton, A.~S. 2011,
  Monthly Notices of the Royal Astronomical Society, 415, 2215

\bibitem[{Bayer {et~al.}(2023{\natexlab{a}})Bayer, Chatterjee, Koopmans,
  Vegetti, McKean, Treu, Fassnacht, Glazebrook, Bayer, Chatterjee, Koopmans,
  Vegetti, McKean, Treu, Fassnacht, \& Glazebrook}]{Bayer2023b}
Bayer, D., Chatterjee, S., Koopmans, L. V.~E., {et~al.} 2023{\natexlab{a}},
  MNRAS, 523, 1310

\bibitem[{Bayer {et~al.}(2023{\natexlab{b}})Bayer, Koopmans, McKean, Vegetti,
  Treu, Fassnacht, \& Glazebrook}]{Bayer2023a}
Bayer, D., Koopmans, L.~V., McKean, J.~P., {et~al.} 2023{\natexlab{b}}, Monthly
  Notices of the Royal Astronomical Society, 523, 1326

\bibitem[{Bell {et~al.}(2003)Bell, McIntosh, Katz, \& Weinberg}]{Bell2003}
Bell, E.~F., McIntosh, D.~H., Katz, N., \& Weinberg, M.~D. 2003, The
  Astrophysical Journal Supplement Series, 149, 289

\bibitem[{Blanton \& Roweis(2007)}]{Blanton2007}
Blanton, M.~R. \& Roweis, S. 2007, The Astronomical Journal, 133, 734

\bibitem[{{Bolton} {et~al.}(2012){Bolton}, {Brownstein}, {Kochanek}, {Shu},
  {Schlegel}, {Eisenstein}, {Wake}, {Connolly}, {Maraston}, {Arneson}, \&
  {Weaver}}]{Bolton2012}
{Bolton}, A.~S., {Brownstein}, J.~R., {Kochanek}, C.~S., {et~al.} 2012, \apj,
  757, 82

\bibitem[{Bolton {et~al.}(2008)Bolton, Burles, Koopmans, Treu, Gavazzi,
  Moustakas, Wayth, \& Schlegel}]{Bolton2008}
Bolton, A.~S., Burles, S., Koopmans, L. V.~E., {et~al.} 2008, The Astrophysical
  Journal, 682, 964

\bibitem[{Bolton {et~al.}(2006)Bolton, Burles, Koopmans, Treu, \&
  Moustakas}]{Bolton2006}
Bolton, A.~S., Burles, S., Koopmans, L. V.~E., Treu, T., \& Moustakas, L.~A.
  2006, The Astrophysical Journal, 638, 703

\bibitem[{Brown {et~al.}(2018)Brown, Vallenari, Prusti, Bruijne, Babusiaux,
  Bailer-Jones, Biermann, Evans, Eyer, Jansen, Jordi, Klioner, Lammers,
  Lindegren, Luri, Mignard, Panem, Pourbaix, Randich, Sartoretti, Siddiqui,
  Soubiran, Leeuwen, Walton, Arenou, Bastian, Cropper, Drimmel, Katz, Lattanzi,
  Bakker, Cacciari, Castaneda, Chaoul, Cheek, Angeli, Fabricius, Guerra, Holl,
  Masana, Messineo, Mowlavi, Nienartowicz, Panuzzo, Portell, Riello, Seabroke,
  Tanga, Thévenin, Gracia-Abril, Comoretto, Garcia-Reinaldos, Teyssier,
  Altmann, Andrae, Audard, Bellas-Velidis, Benson, Berthier, Blomme, Burgess,
  Busso, Carry, Cellino, Clementini, Clotet, Creevey, Davidson, Ridder,
  Delchambre, Dell'Oro, Ducourant, Fernández-Hernández, Fouesneau, Frémat,
  Galluccio, García-Torres, González-Núnez, González-Vidal, Gosset, Guy,
  Halbwachs, Hambly, Harrison, Hernández, Hestroffer, Hodgkin, Hutton,
  Jasniewicz, Jean-Antoine-Piccolo, Jordan, Korn, Krone-Martins, Lanzafame,
  Lebzelter, Löffler, Manteiga, Marrese, Martín-Fleitas, Moitinho, Mora,
  Muinonen, Osinde, Pancino, Pauwels, Petit, Recio-Blanco, Richards, Rimoldini,
  Robin, Sarro, Siopis, Smith, Sozzetti, Süveges, Torra, Reeven, Abbas,
  Aramburu, Accart, Aerts, Altavilla, Álvarez, Alvarez, Alves, Anderson,
  Andrei, Varela, Antiche, Antoja, Arcay, Astraatmadja, Bach, Baker,
  Balaguer-Núnez, Balm, Barache, Barata, Barbato, Barblan, Barklem, Barrado,
  Barros, Barstow, Munoz, Bassilana, Becciani, Bellazzini, Berihuete, Bertone,
  Bianchi, Bienaymé, Blanco-Cuaresma, Boch, Boeche, Bombrun, Borrachero,
  Bossini, Bouquillon, Bourda, Bragaglia, Bramante, Breddels, Bressan,
  Brouillet, Brüsemeister, Brugaletta, Bucciarelli, Burlacu, Busonero,
  Butkevich, Buzzi, Caffau, Cancelliere, Cannizzaro, Cantat-Gaudin, Carballo,
  Carlucci, Carrasco, Casamiquela, Castellani, Castro-Ginard, Charlot, Chemin,
  Chiavassa, Cocozza, Costigan, Cowell, Crifo, Crosta, Crowley, Cuypersy,
  Dafonte, Damerdji, Dapergolas, David, David, Laverny, Luise, March, Martino,
  Souza, Torres, Debosscher, Pozo, Delbo, Delgado, Delgado, Matteo, Diakite,
  Diener, Distefano, Dolding, Drazinos, Durán, Edvardsson, Enke, Eriksson,
  Esquej, Bontemps, Fabre, Fabrizio, Faigler, Falcão, Casas, Federici,
  Fedorets, Fernique, Figueras, Filippi, Findeisen, Fonti, Fraile, Fraser,
  Frézouls, Gai, Galleti, Garabato, García-Sedano, Garofalo, Garralda, Gavel,
  Gavras, Gerssen, Geyer, Giacobbe, Gilmore, Girona, Giuffrida, Glass, Gomes,
  Granvik, Gueguen, Guerrier, Guiraud, Gutiérrez-Sánchez, Haigron,
  Hatzidimitriou, Hauser, Haywood, Heiter, Helmi, Heu, Hilger, Hobbs, Hofmann,
  Holland, Huckle, Hypki, Icardi, Janßen, Fombelle, Jonker, Juhász, Julbe,
  Karampelas, Kewley, Klar, Kochoska, Kohley, Kolenberg, Kontizas, Kontizas,
  Koposov, Kordopatis, Kostrzewa-Rutkowska, Koubsky, Lambert, Lanza, Lasne,
  Lavigne, Fustec, Poncin-Lafitte, Lebreton, Leccia, Leclerc, Lecoeur-Taibi,
  Lenhardt, Leroux, Liao, Licata, Lindstrøm, Lister, Livanou, Lobel, López,
  Managau, Mann, Mantelet, Marchal, Marchant, Marconi, Marinoni, Marschalkó,
  Marshall, Martino, Marton, Mary, Massari, Matijević, Mazeh, McMillan,
  Messina, Michalik, Millar, Molina, Molinaro, Molnár, Montegriffo, Mor,
  Morbidelli, Morel, Morris, Mulone, Muraveva, Musella, Nelemans, Nicastro,
  Noval, O'Mullane, Ordénovic, Ordónez-Blanco, Osborne, Pagani, Pagano,
  Pailler, Palacin, Palaversa, Panahi, Pawlak, Piersimoni, Pineau, Plachy,
  Plum, Poggio, Poujoulet, Prša, Pulone, Racero, Ragaini, Rambaux,
  Ramos-Lerate, Regibo, Reylé, Riclet, Ripepi, Riva, Rivard, Rixon, Roegiers,
  Roelens, Romero-Gómez, Rowell, Royer, Ruiz-Dern, Sadowski, Sellés,
  Sahlmann, Salgado, Salguero, Sanna, Santana-Ros, Sarasso, Savietto,
  Schultheis, Sciacca, Segol, Segovia, Ségransan, Shih, Siltala, Silva, Smart,
  Smith, Solano, Solitro, Sordo, Nieto, Souchay, Spagna, Spoto, Stampa, Steele,
  Steidelmüller, Stephenson, Stoev, Suess, Surdej, Szabados, Szegedi-Elek,
  Tapiador, Taris, Tauran, Taylor, Teixeira, Terrett, Teyssandier, Thuillot,
  Titarenko, Clotet, Turon, Ulla, Utrilla, Uzzi, Vaillant, Valentini, Valette,
  Elteren, Hemelryck, Leeuwen, Vaschetto, Vecchiato, Veljanoski, Viala,
  Vicente, Vogt, Essen, Voss, Votruba, Voutsinas, Walmsley, Weiler, Wertz,
  Wevers, Wyrzykowski, Yoldas, Žerjal, Ziaeepour, Zorec, Zschocke, Zucker,
  Zurbach, \& Zwitter}]{Brown2018}
Brown, A.~G., Vallenari, A., Prusti, T., {et~al.} 2018, A \& A, 616, A1

\bibitem[{Bunker(2019)}]{Bunker2019}
Bunker, A.~J. 2019, Proceedings of the International Astronomical Union, 15,
  342

\bibitem[{Cappellari(2016)}]{Cappellari2016}
Cappellari, M. 2016, https://doi.org/10.1146/annurev-astro-082214-122432, 54,
  597

\bibitem[{Cohn {et~al.}(2001)Cohn, Kochanek, McLeod, \& Keeton}]{Cohn2001}
Cohn, J.~D., Kochanek, C.~S., McLeod, B.~A., \& Keeton, C.~R. 2001, The
  Astrophysical Journal, 554, 1216

\bibitem[{Cole {et~al.}(2000)Cole, Lacey, Baugh, \& Frenk}]{Cole2000}
Cole, S., Lacey, C.~G., Baugh, C.~M., \& Frenk, C.~S. 2000, Monthly Notices of
  the Royal Astronomical Society, 319, 168

\bibitem[{Dalal \& Kochanek(2001)}]{Dalal2001}
Dalal, N. \& Kochanek, C.~S. 2001, ApJ, 572, 25

\bibitem[{Despali {et~al.}(2020)Despali, Lovell, Vegetti, Crain, \&
  Oppenheimer}]{Despali2020}
Despali, G., Lovell, M., Vegetti, S., Crain, R.~A., \& Oppenheimer, B.~D. 2020,
  Monthly Notices of the Royal Astronomical Society, 491, 1295

\bibitem[{Despali {et~al.}(2017)Despali, Vegetti, White, \&
  Giocoli}]{Despali2017}
Despali, G., Vegetti, S., White, S. D.~M., \& Giocoli, C. 2017, Mon. Not. R.
  Astron. Soc, 000

\bibitem[{Diemand {et~al.}(2008)Diemand, Kuhlen, Madau, Zemp, Moore, Potter, \&
  Stadel}]{Diemand2008}
Diemand, J., Kuhlen, M., Madau, P., {et~al.} 2008, Nature 2008 454:7205, 454,
  735

\bibitem[{Dullo \& Graham(2014)}]{Dullo2014}
Dullo, B.~T. \& Graham, A.~W. 2014, MNRAS, 444, 2700

\bibitem[{Eigenbrod {et~al.}(2006)Eigenbrod, Courbin, Meylan, Vuissoz, \&
  Magain}]{Eigenbrod2006}
Eigenbrod, A., Courbin, F., Meylan, G., Vuissoz, C., \& Magain, P. 2006, A\&A,
  451, 759

\bibitem[{Ertl {et~al.}(2023)Ertl, Schuldt, Suyu, Schmidt, Treu, Birrer,
  Shajib, \& Sluse}]{Ertl2023}
Ertl, S., Schuldt, S., Suyu, S.~H., {et~al.} 2023, A\&A, 672, A2

\bibitem[{Etherington {et~al.}(2023)Etherington, Nightingale, Massey, Tam, Cao,
  Niemiec, He, Robertson, Li, Amvrosiadis, Cole, Diego, Frenk, Frye, Harvey,
  Jauzac, Koekemoer, Lagattuta, Limousin, Mahler, Sirks, \&
  Steinhardt}]{Etherington2023}
Etherington, A., Nightingale, J.~W., Massey, R., {et~al.} 2023, MNRAS, 000, 1

\bibitem[{Faber {et~al.}(1997)Faber, Tremaine, Ajhar, Byun, Dressler, Gebhardt,
  Grillmair, Kormendy, Lauer, \& Richstone}]{Faber1997}
Faber, S.~M., Tremaine, S., Ajhar, E.~A., {et~al.} 1997, The Astronomical
  Journal, 114, 1771

\bibitem[{Falco {et~al.}(1999)Falco, Kochanek, Lehar, McLeod, Munoz, Impey,
  Keeton, Peng, \& Rix}]{Falco1999}
Falco, E.~E., Kochanek, C.~S., Lehar, J., {et~al.} 1999

\bibitem[{Faure {et~al.}(2004)Faure, Alloin, Kneib, \& Courbin}]{Faure2004}
Faure, C., Alloin, D., Kneib, J.~P., \& Courbin, F. 2004, Astronomy \&
  Astrophysics, 428, 741

\bibitem[{Finkelstein {et~al.}(2022)Finkelstein, Bagley, Ferguson, Wilkins,
  Kartaltepe, Papovich, Yung, Haro, Behroozi, Dickinson, Kocevski, Koekemoer,
  Larson, Bail, Morales, Perez-Gonzalez, Burgarella, Dave, Hirschmann,
  Somerville, Wuyts, Bromm, Casey, Fontana, Fujimoto, Gardner, Giavalisco,
  Grazian, Grogin, Hathi, Hutchison, Jha, Jogee, Kewley, Kirkpatrick, Long,
  Lotz, Pentericci, Pierel, Pirzkal, Ravindranath, Ryan, Trump, Yang,
  Bhatawdekar, Bisigello, Buat, Calabro, Castellano, Cleri, Cooper, Croton,
  Daddi, Dekel, Elbaz, Franco, Gawiser, Holwerda, Huertas-Company, Jaskot,
  Leung, Lucas, Mobasher, Pandya, Tacchella, Weiner, \&
  Zavala}]{Finkelstein2022}
Finkelstein, S.~L., Bagley, M.~B., Ferguson, H.~C., {et~al.} 2022, ApJL, 946,
  L13

\bibitem[{Gavazzi {et~al.}(2012)Gavazzi, Treu, Marshall, Brault, \&
  Ruff}]{Gavazzi2012}
Gavazzi, R., Treu, T., Marshall, P.~J., Brault, F., \& Ruff, A. 2012,
  Astrophysical Journal, 761

\bibitem[{Gilman {et~al.}(2019)Gilman, Birrer, Treu, Nierenberg, \&
  Benson}]{Gilman2019}
Gilman, D., Birrer, S., Treu, T., Nierenberg, A., \& Benson, A. 2019, Monthly
  Notices of the Royal Astronomical Society, 487, 5721

\bibitem[{Gomer(2020)}]{GomerTh2020}
Gomer, M.~R. 2020, Retrieved from the University of Minnesota Digital
  Conservancy, hdl.handle.net/11299/216394

\bibitem[{Graham(2005)}]{Graham2005}
Graham, A.~W. 2005, ApJL, 613, L33

\bibitem[{Harrison {et~al.}(1974)Harrison, Harrison, \& R.}]{Harrison1974}
Harrison, E.~R., Harrison, \& R., E. 1974, ApJL, 191, L51

\bibitem[{Hogg {et~al.}(2002)Hogg, Baldry, Blanton, \& Eisenstein}]{Hogg2002}
Hogg, D.~W., Baldry, I.~K., Blanton, M.~R., \& Eisenstein, D.~J. 2002

\bibitem[{Hsueh {et~al.}(2020)Hsueh, Enzi, Vegetti, Auger, Fassnacht, Despali,
  Koopmans, \& McKean}]{Hsueh2020}
Hsueh, J.~W., Enzi, W., Vegetti, S., {et~al.} 2020, Monthly Notices of the
  Royal Astronomical Society, 492, 3047

\bibitem[{Hunter(2007)}]{Hunter2007}
Hunter, J.~D. 2007, Computing in Science and Engineering, 9, 90

\bibitem[{Jackson {et~al.}(2010)Jackson, Bryan, Mao, \& Li}]{Jackson2010}
Jackson, N., Bryan, S.~E., Mao, S., \& Li, C. 2010, Monthly Notices of the
  Royal Astronomical Society, 403, 826

\bibitem[{Jones {et~al.}(2001)Jones, Oliphant, \& Peterson}]{Jones2001}
Jones, E., Oliphant, T., \& Peterson, P. 2001

\bibitem[{Kauffmann {et~al.}(2003)Kauffmann, Heckman, White, Charlot, Tremonti,
  Brinchmann, Bruzual, Peng, Seibert, Bernardi, Blanton, Brinkmann, Castander,
  Csábai, Fukugita, Ivezic, Munn, Nichol, Padmanabhan, Thakar, Weinberg, \&
  York}]{Kauffmann2003}
Kauffmann, G., Heckman, T.~M., White, S. D.~M., {et~al.} 2003, Mon. Not. R.
  Astron. Soc, 341, 33

\bibitem[{Kauffmann {et~al.}(1993)Kauffmann, White, Guiderdoni, Kauffmann,
  White, \& Guiderdoni}]{Kauffmann1993}
Kauffmann, G., White, S. D.~M., Guiderdoni, B., {et~al.} 1993, MNRAS, 264, 201

\bibitem[{Keeton {et~al.}(1997)Keeton, Kochanek, \& Seljak}]{Keeton1997}
Keeton, C.~R., Kochanek, C.~S., \& Seljak, U. 1997, The Astrophysical Journal,
  482, 604

\bibitem[{King {et~al.}(1966)King, Minkowski, King, \& Minkowski}]{King1966}
King, I.~R., Minkowski, R., King, I.~R., \& Minkowski, R. 1966, ApJ, 143, 1002

\bibitem[{Kochanek {et~al.}(1991)Kochanek, Kochanek, \& S.}]{Kochanek1991}
Kochanek, C.~S., Kochanek, \& S., C. 1991, ApJ, 373, 354

\bibitem[{Koopmans(2005)}]{Koopmans2005}
Koopmans, L.~V. 2005, Monthly Notices of the Royal Astronomical Society, 363,
  1136

\bibitem[{Kormendy \& Bender(2009)}]{Kormendy2009}
Kormendy, J. \& Bender, R. 2009, The Astrophysical Journal, 691, 142

\bibitem[{Labbé {et~al.}(2023)Labbé, van Dokkum, Nelson, Bezanson, Suess,
  Leja, Brammer, Whitaker, Mathews, Stefanon, \& Wang}]{Labbe2023}
Labbé, I., van Dokkum, P., Nelson, E., {et~al.} 2023, Nature 2023 616:7956,
  616, 266

\bibitem[{Lauer {et~al.}(1995)Lauer, Ajhar, Byun, Dressler, Faber, Grillmair,
  Kormendy, Richstone, Tremaine, Lauer, Ajhar, Byun, Dressler, Faber,
  Grillmair, Kormendy, Richstone, \& Tremaine}]{Lauer1995}
Lauer, T.~R., Ajhar, E.~A., Byun, Y.~I., {et~al.} 1995, AJ, 110, 2622

\bibitem[{Lemon {et~al.}(2018)Lemon, Auger, McMahon, \& Ostrovski}]{Lemon2018}
Lemon, C.~A., Auger, M.~W., McMahon, R.~G., \& Ostrovski, F. 2018, Monthly
  Notices of the Royal Astronomical Society, 479, 5060

\bibitem[{Lin {et~al.}(2017)Lin, Buckley-Geer, Agnello, Ostrovski, McMahon,
  Nord, Kuropatkin, Tucker, Treu, Chan, Suyu, Diehl, Collett, Gill, More,
  Amara, Auger, Courbin, Fassnacht, Frieman, Marshall, Meylan, Rusu, Abbott,
  Abdalla, Allam, Banerji, Bechtol, Benoit-Lévy, Bertin, Brooks, Burke,
  Rosell, Kind, Carretero, Castander, Crocce, D'Andrea, da~Costa, Desai,
  Dietrich, Eifler, Finley, Flaugher, Fosalba, García-Bellido, Gaztanaga,
  Gerdes, Goldstein, Gruen, Gruendl, Gschwend, Gutierrez, Honscheid, James,
  Kuehn, Lahav, Li, Lima, Maia, March, Marshall, Martini, Melchior, Menanteau,
  Miquel, Ogando, Plazas, Romer, Sanchez, Schindler, Schubnell, Sevilla-Noarbe,
  Smith, Smith, Sobreira, Suchyta, Swanson, Tarle, Thomas, \& Walker}]{Lin2017}
Lin, H., Buckley-Geer, E., Agnello, A., {et~al.} 2017, The Astrophysical
  Journal, 838, L15

\bibitem[{Lucey {et~al.}(2018)Lucey, Schechter, Smith, \& Anguita}]{Lucey2018}
Lucey, J.~R., Schechter, P.~L., Smith, R.~J., \& Anguita, T. 2018, Monthly
  Notices of the Royal Astronomical Society, 476, 927

\bibitem[{Luhtaru {et~al.}(2021)Luhtaru, Schechter, \& de~Soto}]{Luhtaru2021}
Luhtaru, R., Schechter, P.~L., \& de~Soto, K.~M. 2021, The Astrophysical
  Journal, 915, 4

\bibitem[{Mao \& Schneider(1997)}]{Mao1997}
Mao, S. \& Schneider, P. 1997, Monthly Notices of the Royal Astronomical
  Society, 295, 587

\bibitem[{Mccaffrey {et~al.}(2023)Mccaffrey, Hardin, Wise, \&
  Regan}]{Mccaffrey2023}
Mccaffrey, J.~M., Hardin, S.~E., Wise, J.~H., \& Regan, J.~A. 2023

\bibitem[{Metcalf \& Madau(2001)}]{Metcalf2001}
Metcalf, R.~B. \& Madau, P. 2001, The Astrophysical Journal, 563, 9

\bibitem[{Millon {et~al.}(2020)Millon, Courbin, Bonvin, Paic, Meylan, Tewes,
  Sluse, Magain, Chan, Galan, Joseph, Lemon, Tihhonova, Anderson, Marmier,
  Chazelas, Lendl, Triaud, \& Wyttenbach}]{Millon2020}
Millon, M., Courbin, F., Bonvin, V., {et~al.} 2020, Astronomy \& Astrophysics,
  640, A105

\bibitem[{Milosavljevi\'{c} {et~al.}(2002)Milosavljevi\'{c}, Merritt, Rest, \&
  Bosch}]{Milosavljevic2002}
Milosavljevi\'{c}, M., Merritt, D., Rest, A., \& Bosch, F. C. V.~D. 2002,
  Monthly Notices of the Royal Astronomical Society, 331, L51

\bibitem[{Minor {et~al.}(2017)Minor, Kaplinghat, \& Li}]{Minor2017}
Minor, Q.~E., Kaplinghat, M., \& Li, N. 2017, The Astrophysical Journal, 845,
  118

\bibitem[{Myers {et~al.}(1995)Myers, Fassnacht, Djorgovski, Blandford,
  Matthews, Neugebauer, Pearson, Readhead, Smith, Thompson, Womble, Browne,
  Wilkinson, Nair, Jackson, Snellen, Miley, de~Bruyn, Schilizzi, Myers,
  Fassnacht, Djorgovski, Blandford, Matthews, Neugebauer, Pearson, Readhead,
  Smith, Thompson, Womble, Browne, Wilkinson, Nair, Jackson, Snellen, Miley,
  de~Bruyn, \& Schilizzi}]{Myers1995}
Myers, S.~T., Fassnacht, C.~D., Djorgovski, S.~G., {et~al.} 1995, ApJL, 447, L5

\bibitem[{Nasim {et~al.}(2021)Nasim, Gualandris, Read, Antonini, Dehnen, \&
  Delorme}]{Nasim2021}
Nasim, I.~T., Gualandris, A., Read, J.~I., {et~al.} 2021, Monthly Notices of
  the Royal Astronomical Society, 502, 4794

\bibitem[{Navarro {et~al.}(1996)Navarro, Frenk, \& White}]{Navarro1996}
Navarro, J.~F., Frenk, C.~S., \& White, S. D.~M. 1996, The Astrophysical
  Journal, 462, 563

\bibitem[{Navarro {et~al.}(1997)Navarro, Frenk, \& White}]{Navarro1997}
Navarro, J.~F., Frenk, C.~S., \& White, S. D.~M. 1997, The Astrophysical
  Journal, 490, 493

\bibitem[{Nierenberg {et~al.}(2017)Nierenberg, Treu, Brammer, Peter, Fassnacht,
  Keeton, Kochanek, Schmidt, Sluse, \& Wright}]{Nierenberg2017}
Nierenberg, A.~M., Treu, T., Brammer, G., {et~al.} 2017, Monthly Notices of the
  Royal Astronomical Society, 471, 2224

\bibitem[{Nierenberg {et~al.}(2014)Nierenberg, Treu, Wright, Fassnacht, \&
  Auger}]{Nierenberg2014}
Nierenberg, A.~M., Treu, T., Wright, S.~A., Fassnacht, C.~D., \& Auger, M.~W.
  2014, Monthly Notices of the Royal Astronomical Society, 442, 2434

\bibitem[{Oliphant(2015)}]{Oliphant2015}
Oliphant, T.~E. 2015

\bibitem[{Ostrovski {et~al.}(2018)Ostrovski, Lemon, Auger, McMahon, Fassnacht,
  Chen, Connolly, Koposov, Pons, Reed, \& Rusu}]{Ostrovski2018}
Ostrovski, F., Lemon, C.~A., Auger, M.~W., {et~al.} 2018, Monthly Notices of
  the Royal Astronomical Society: Letters, 473, L116

\bibitem[{Postman {et~al.}(2012)Postman, Lauer, Donahue, Graves, Coe,
  Moustakas, Koekemoer, Bradley, Ford, Grillo, Zitrin, Lemze, Broadhurst,
  Moustakas, Ascaso, Medezinski, \& Kelson}]{Postman2012}
Postman, M., Lauer, T.~R., Donahue, M., {et~al.} 2012, ApJ, 756, 159

\bibitem[{Ritondale {et~al.}(2019)Ritondale, Vegetti, Despali, Auger, Koopmans,
  \& Mckean}]{Ritondale2019}
Ritondale, E., Vegetti, S., Despali, G., {et~al.} 2019, MNRAS, 485, 2179

\bibitem[{Rusin \& Ma(2000)}]{Rusin2000}
Rusin, D. \& Ma, C.-P. 2000, The Astrophysical Journal, 549, L33

\bibitem[{Rusli {et~al.}(2013)Rusli, Erwin, Saglia, Thomas, Fabricius, Bender,
  \& Nowak}]{Rusli2013}
Rusli, S.~P., Erwin, P., Saglia, R.~P., {et~al.} 2013, The Astronomical
  Journal, 146, 160

\bibitem[{Ryden {et~al.}(1992)Ryden, Ryden, \& Barbara}]{Ryden1992}
Ryden, B., Ryden, \& Barbara. 1992, ApJ, 396, 445

\bibitem[{Sahu {et~al.}(in prep.)Sahu, Tran, Suyu, Shajib, Ertl, Kacprzak,
  Glazebrook, Jones, G.C., Barone, Baker, Skobe, Derkenne, Lewis, Sweet, \&
  Lopez}]{Sahu2023}
Sahu, N., Tran, K., Suyu, S.~H., {et~al.} in prep.

\bibitem[{Schuldt {et~al.}(2023)Schuldt, Suyu, Canameras, Shu, Taubenberger,
  Ertl, \& Halkola}]{Schuldt2022}
Schuldt, S., Suyu, S.~H., Canameras, R., {et~al.} 2023, Astronomy \&
  Astrophysics, 673, A33

\bibitem[{S{\'e}rsic(1963)}]{Sersic.1963}
S{\'e}rsic, J. 1963, Boletin de la Asociacion Argentina de Astronomia La Plata
  Argentina

\bibitem[{Shajib {et~al.}(2021)Shajib, Treu, Birrer, \&
  Sonnenfeld}]{Shajib2021}
Shajib, A.~J., Treu, T., Birrer, S., \& Sonnenfeld, A. 2021, Monthly Notices of
  the Royal Astronomical Society, 503, 2380

\bibitem[{Simon \& Geha(2007)}]{Simon2007}
Simon, J.~D. \& Geha, M. 2007, The Astrophysical Journal, 670, 313

\bibitem[{Sonnenfeld {et~al.}(2011)Sonnenfeld, Treu, Gavazzi, Marshall, Auger,
  Suyu, Koopmans, \& Bolton}]{Sonnenfeld2011}
Sonnenfeld, A., Treu, T., Gavazzi, R., {et~al.} 2011, ApJ, 752, 163

\bibitem[{Sonnenfeld {et~al.}(2015)Sonnenfeld, Treu, Marshall, Suyu, Gavazzi,
  Auger, \& Nipoti}]{Sonnenfeld2015}
Sonnenfeld, A., Treu, T., Marshall, P.~J., {et~al.} 2015, Astrophysical
  Journal, 800, 94

\bibitem[{Suyu \& Halkola(2010)}]{Suyu.2010}
Suyu, S.~H. \& Halkola, A. 2010, Astronomy {\&} Astrophysics, 524, A94

\bibitem[{Suyu {et~al.}(2012)Suyu, Hensel, McKean, Fassnacht, Treu, Halkola,
  Norbury, Jackson, Schneider, Thompson, Auger, Koopmans, \&
  Matthews}]{Suyu.2012}
Suyu, S.~H., Hensel, S.~W., McKean, J.~P., {et~al.} 2012, The Astrophysical
  Journal, 750, 10

\bibitem[{Suyu {et~al.}(2009)Suyu, Marshall, Auger, Hilbert, Blandford,
  Koopmans, Fassnacht, \& Treu}]{Suyu2009}
Suyu, S.~H., Marshall, P.~J., Auger, M.~W., {et~al.} 2009, Astrophysical
  Journal, 711, 201

\bibitem[{{The Astropy Collaboration} {et~al.}(2018){The Astropy
  Collaboration}, Price-Whelan, Sipőcz, Günther, Lim, Crawford, Conseil,
  Shupe, Craig, Dencheva, Ginsburg, VanderPlas, Bradley, Pérez-Suárez,
  de~Val-Borro, Contributors), Aldcroft, Cruz, Robitaille, Tollerud,
  Committee), Ardelean, Babej, Bach, Bachetti, Bakanov, Bamford, Barentsen,
  Barmby, Baumbach, Berry, Biscani, Boquien, Bostroem, Bouma, Brammer, Bray,
  Breytenbach, Buddelmeijer, Burke, Calderone, Rodríguez, Cara, Cardoso,
  Cheedella, Copin, Corrales, Crichton, D’Avella, Deil, Depagne, Dietrich,
  Donath, Droettboom, Earl, Erben, Fabbro, Ferreira, Finethy, Fox, Garrison,
  Gibbons, Goldstein, Gommers, Greco, Greenfield, Groener, Grollier, Hagen,
  Hirst, Homeier, Horton, Hosseinzadeh, Hu, Hunkeler, Ivezić, Jain, Jenness,
  Kanarek, Kendrew, Kern, Kerzendorf, Khvalko, King, Kirkby, Kulkarni, Kumar,
  Lee, Lenz, Littlefair, Ma, Macleod, Mastropietro, McCully, Montagnac, Morris,
  Mueller, Mumford, Muna, Murphy, Nelson, Nguyen, Ninan, Nöthe, Ogaz, Oh,
  Parejko, Parley, Pascual, Patil, Patil, Plunkett, Prochaska, Rastogi, Janga,
  Sabater, Sakurikar, Seifert, Sherbert, Sherwood-Taylor, Shih, Sick, Silbiger,
  Singanamalla, Singer, Sladen, Sooley, Sornarajah, Streicher, Teuben, Thomas,
  Tremblay, Turner, Terrón, van Kerkwijk, de~la Vega, Watkins, Weaver,
  Whitmore, Woillez, Zabalza, \& Contributors)}]{AstropyCollaboration2018}
{The Astropy Collaboration}, Price-Whelan, A.~M., Sipőcz, B.~M., {et~al.}
  2018, The Astronomical Journal, 156, 123

\bibitem[{Toomre \& Toomre(1972)}]{Toomre1972}
Toomre, A. \& Toomre, J. 1972, The Astrophysical Journal, 178, 623

\bibitem[{Vegetti {et~al.}(2014)Vegetti, Koopmans, Auger, Treu, \&
  Bolton}]{Vegetti2014b}
Vegetti, S., Koopmans, L.~V., Auger, M.~W., Treu, T., \& Bolton, A.~S. 2014,
  Monthly Notices of the Royal Astronomical Society, 442, 2017

\bibitem[{Vegetti {et~al.}(2010)Vegetti, Koopmans, Bolton, Treu, \&
  Gavazzi}]{Vegetti2010}
Vegetti, S., Koopmans, L.~V., Bolton, A., Treu, T., \& Gavazzi, R. 2010,
  Monthly Notices of the Royal Astronomical Society, 408, 1969

\bibitem[{Vegetti {et~al.}(2012)Vegetti, Lagattuta, McKean, Auger, Fassnacht,
  \& Koopmans}]{Vegetti2012}
Vegetti, S., Lagattuta, D.~J., McKean, J.~P., {et~al.} 2012, Nature 2012
  481:7381, 481, 341

\bibitem[{Vegetti \& Vogelsberger(2014)}]{Vegetti2014}
Vegetti, S. \& Vogelsberger, M. 2014, Monthly Notices of the Royal Astronomical
  Society, 442, 3598

\bibitem[{Wagner(2017)}]{Wagner2017}
Wagner, J. 2017, Astronomy and Astrophysics, 615

\bibitem[{Wagner(2019)}]{Wagner2019}
Wagner, J. 2019, Universe, 5, 177

\bibitem[{Wagner(2020)}]{Wagner2020}
Wagner, J. 2020, General Relativity and Gravitation, 52, 61

\bibitem[{Weisenbach {et~al.}(2021)Weisenbach, Schechter, \&
  Pontula}]{Weisenbach2021}
Weisenbach, L., Schechter, P.~L., \& Pontula, S. 2021, The Astrophysical
  Journal, 922, 70

\bibitem[{White {et~al.}(1991)White, Frenk, White, \& Frenk}]{White1991}
White, S. D.~M., Frenk, C.~S., White, S. D.~M., \& Frenk, C.~S. 1991, ApJ, 379,
  52

\bibitem[{Wisotzki {et~al.}(1999)Wisotzki, Christlieb, Liu, Maza, Morgan,
  Schechter, Wisotzki, Christlieb, Liu, Maza, Morgan, \&
  Schechter}]{Wisotzki1999}
Wisotzki, L., Christlieb, N., Liu, M.~C., {et~al.} 1999, A\&A, 348, L41

\bibitem[{Wisotzki {et~al.}(1996)Wisotzki, Koehler, Groote, Reimers, Wisotzki,
  Koehler, Groote, \& Reimers}]{Wisotzki1996}
Wisotzki, L., Koehler, T., Groote, D., {et~al.} 1996, A\&A, 115, 227

\bibitem[{Wong {et~al.}(2011)Wong, Keeton, Williams, Momcheva, \&
  Zabludoff}]{Wong2011}
Wong, K.~C., Keeton, C.~R., Williams, K.~A., Momcheva, I.~G., \& Zabludoff,
  A.~I. 2011, The Astrophysical Journal, 726, 84

\bibitem[{Xivry \& Marshall(2008)}]{Orban2008}
Xivry, G. O.~D. \& Marshall, P. 2008, Mon. Not. R. Astron. Soc, 000, 0

\bibitem[{Xu {et~al.}(2014)Xu, Sluse, Gao, Wang, Frenk, Mao, Schneider, \&
  Springel}]{Xu2014}
Xu, D., Sluse, D., Gao, L., {et~al.} 2014, Monthly Notices of the Royal
  Astronomical Society, 447, 3189

\end{thebibliography}

\clearpage

\begin{appendix}

\begin{landscape}

\section{Final model parameter values of all candidate model classes}

\begin{table}[h]
\caption{Final model parameter values of all candidate model classes.}

\newcolumntype{L}{>{\raggedright\arraybackslash}X}
\fontsize{7}{8}\selectfont
\begin{tabularx}{\linewidth}{lllLLLLLL}\toprule \toprule
        Parameter description & Parameter &  & \multicolumn{6}{c}{Model class} \\ \cmidrule(lr){4-9}
        & & & PL1 + cPL2 + $\gamma_{\rm ext}$ & cPL1 + cPL2 + $\gamma_{\rm ext}$ & PL1 + PL2 + rSIS & PL1 + PL2 + rNIS & PL1 + PL2 + rSIS + $\gamma_{\rm ext}$ & PL1 + PL2 + rNIS + $\gamma_{\rm ext}$ \\
        \toprule \toprule  
        \multicolumn{2}{l}{G1 lens mass (power-law)} & & & & \\ \midrule
        \ \\
        axis-ratio & $q_{\rm G1}$ & full chain &  ${0.78}^{+0.05}_{-0.06} $ & ${0.85}^{+0.03}_{-0.04} $ & ${0.79}^{+0.03}_{-0.04} $& ${0.80}^{+0.03}_{-0.04} $ & ${0.83}^{+0.04}_{-0.05} $ & ${0.86}^{+0.06}_{-0.05} $   \vspace{4px} \\ 
         &  & candidates &  ${0.87}^{+0.04}_{-0.06} $ & ${0.90}^{+0.03}_{-0.03} $ & ${0.84}^{+0.01}_{-0.02} $ & ${0.83}^{+0.02}_{-0.05} $ & ${0.90}^{+0.03}_{-0.03} $ & ${0.91}^{+0.04}_{-0.04} $     \vspace{10px} \\ 
         
          position angle & $\phi_{\rm G1}$ [\degree] & full chain &  ${118}^{+8}_{-7} $ & ${109}^{+8}_{-8} $ & ${107}^{+13}_{-11} $ & ${91}^{+7}_{-7} $ & ${104}^{+10}_{-10} $ & ${101}^{+10}_{-10} $     \vspace{4px} \\ 
         &  & candidates &  ${122}^{+12}_{-9} $ & ${113}^{+9}_{-9} $ & ${112}^{+4}_{-3} $ & ${89}^{+7}_{-8} $ & ${104}^{+9}_{-9} $ & ${101}^{+11}_{-10} $     \vspace{10px} \\     
         
          Einstein radius & $\theta_{\rm E,G1}\ ['']$ & full chain &  ${0.89}^{+0.04}_{-0.04} $ & ${0.85}^{+0.04}_{-0.05} $ & ${0.81}^{+0.04}_{-0.04} $ & ${0.75}^{+0.04}_{-0.04} $ & ${0.86}^{+0.04}_{-0.04} $ & ${0.80}^{+0.06}_{-0.07} $     \vspace{4px} \\ 
         &  & candidates &  ${0.85}^{+0.03}_{-0.04} $  & ${0.85}^{+0.03}_{-0.03} $ & ${0.76}^{+0.02}_{-0.03} $ & ${0.70}^{+0.05}_{-0.06} $  & ${0.81}^{+0.02}_{-0.03} $ & ${0.77}^{+0.04}_{-0.05} $     \vspace{10px} \\  
         
        core radius & $r_{\rm c,G1}\ ['']$ & full chain &  $-$ & ${0.25}^{+0.10}_{-0.13} $ & $-$ & $-$ & $-$ & $-$     \vspace{4px} \\ 
         &  & candidates &  $-$ & ${0.16}^{+0.04}_{-0.06} $ & $-$ & $-$ & $-$ & $-$        \vspace{10px} \\  

          power-law index & $\gamma_{\rm G1}$  & full chain &  ${1.97}^{+0.14}_{-0.15} $ & ${2.07}^{+0.16}_{-0.15} $ & ${1.96}^{+0.23}_{-0.21} $ & ${2.17}^{+0.14}_{-0.14} $ & ${2.14}^{+0.15}_{-0.15} $ & ${2.19}^{+0.14}_{-0.14} $     \vspace{4px} \\ 
         &  & candidates &  ${1.88}^{+0.14}_{-0.15} $ & ${2.01}^{+0.16}_{-0.14} $ & ${1.66}^{+0.08}_{-0.07} $ & ${2.13}^{+0.20}_{-0.27} $ & ${2.04}^{+0.14}_{-0.15} $ &  ${2.13}^{+0.14}_{-0.13} $  \\  

  \\
        \ \\ \midrule
                \multicolumn{2}{l}{G2 lens mass (power-law)} & & & & \\ \midrule
        \ \\
         axis-ratio & $q_{\rm G2}$ & full chain &  ${0.77}^{+0.15}_{-0.12} $ & ${0.80}^{+0.14}_{-0.14} $ & ${0.81}^{+0.13}_{-0.14} $ & ${0.80}^{+0.14}_{-0.14} $ & ${0.79}^{+0.14}_{-0.13} $ & ${0.80}^{+0.14}_{-0.14} $     \vspace{4px} \\ 
         &  & candidates &  ${0.97}^{+0.02}_{-0.05} $ & ${0.97}^{+0.03}_{-0.05} $ & ${0.96}^{+0.03}_{-0.07} $ & ${0.75}^{+0.20}_{-0.12} $ & ${0.94}^{+0.05}_{-0.09} $ & ${0.86}^{+0.10}_{-0.15} $      \vspace{10px} \\ 
         
          position angle & $\phi_{\rm G2}$ [\degree] & full chain &  ${50}^{+10}_{-10} $ & ${51}^{+10}_{-10} $ & ${52}^{+10}_{-10} $ & ${52.0}^{+10}_{-10} $ & ${52}^{+9}_{-9} $ & ${52}^{+10}_{-10} $     \vspace{4px} \\ 
         &  & candidates &  ${53}^{+8}_{-10} $ & ${54}^{+11}_{-11} $ & ${55}^{+10}_{-10} $ & ${52}^{+9}_{-10} $ & ${54}^{+10}_{-10} $ & ${53}^{+10}_{-10} $     \vspace{10px} \\     
         
          Einstein radius & $\theta_{\rm E,G2}\ ['']$ & full chain &  ${0.42}^{+0.07}_{-0.07} $ & ${0.32}^{+0.08}_{-0.08} $ & ${0.34}^{+0.08}_{-0.07} $ & ${0.33}^{+0.07}_{-0.07} $ & ${0.36}^{+0.07}_{-0.09} $ & ${0.31}^{+0.09}_{-0.10} $     \vspace{4px} \\ 
         &  & candidates &  ${0.3}^{+0.07}_{-0.08} $ & ${0.28}^{+0.05}_{-0.05} $ & ${0.20}^{+0.02}_{-0.03} $ & ${0.33}^{+0.11}_{-0.14} $ & ${0.29}^{+0.04}_{-0.05} $ & ${0.22}^{+0.06}_{-0.06} $     \vspace{10px} \\  
         
        core radius & $r_{\rm c,G2}\ ['']$ & full chain & ${0.04}^{+0.05}_{-0.03} $  & ${0.06}^{+0.07}_{-0.04} $ & $-$ & $-$ & $-$ & $-$     \vspace{4px} \\ 
         &  & candidates & ${0.14}^{+0.04}_{-0.04} $   & ${0.13}^{+0.04}_{-0.04} $ & $-$ & $-$ & $-$ & $-$        \vspace{10px} \\    

          power-law index & $\gamma_{\rm G2}$  & full chain &  ${2.26}^{+0.14}_{-0.13} $ & ${2.18}^{+0.14}_{-0.14} $ & ${2.01}^{+0.12}_{-0.11} $ &  ${2.16}^{+0.15}_{-0.14} $ & ${2.12}^{+0.15}_{-0.15} $ & ${2.12}^{+0.15}_{-0.14} $     \vspace{4px} \\ 
         &  & candidates &  ${2.26}^{+0.17}_{-0.14} $ & ${2.23}^{+0.14}_{-0.14} $ & ${1.89}^{+0.06}_{-0.08} $ & ${2.01}^{+0.13}_{-0.1} $ & ${1.94}^{+0.07}_{-0.08} $ & ${2.02}^{+0.12}_{-0.11} $  \\  
  \\
        \ \\ \midrule

        \end{tabularx}

        \label{tab:results_param}
        \end{table}
        
\clearpage
        
      \begin{table}[h]
 \captionsetup{labelformat=empty} 
        \caption{Table \ref{tab:results_param} continued.}

\newcolumntype{L}{>{\raggedright\arraybackslash}X}
\fontsize{7}{8}\selectfont
\begin{tabularx}{\linewidth}{lllLLLLLL}\toprule \toprule
        Parameter description & Parameter &  & \multicolumn{6}{c}{Model class} \\ \cmidrule(lr){4-9}
        & & & PL1 + cPL2 + $\gamma_{\rm ext}$ & cPL1 + cPL2 + $\gamma_{\rm ext}$ & PL1 + PL2 + rSIS & PL1 + PL2 + rNIS & PL1 + PL2 + rSIS + $\gamma_{\rm ext}$ & PL1 + PL2 + rNIS + $\gamma_{\rm ext}$ \\
        \toprule \toprule  
                \multicolumn{2}{l}{satellite mass (rSIS/rNIS)} & & & & \\ \midrule
        \ \\

        $x$-centroid & $x_{\rm sat}\ ['']$  & full chain &  $-$ & $-$ & ${0.84}^{+0.17}_{-0.16} $ & ${0.83}^{+0.12}_{-0.13} $ & ${0.71}^{+0.14}_{-0.14} $ & ${0.71}^{+0.13}_{-0.14} $     \vspace{4px} \\ 
         &  & candidates &  $-$ & $-$ & ${0.79}^{+0.14}_{-0.19} $ & ${0.84}^{+0.14}_{-0.12} $ & ${0.77}^{+0.11}_{-0.11} $ & ${0.69}^{+0.11}_{-0.12} $   \vspace{10px}\\ 
        
        $y$-centroid & $y_{\rm sat}\ ['']$  & full chain &  $-$ & $-$ & ${2.22}^{+0.15}_{-0.14} $ & ${2.12}^{+0.13}_{-0.13} $ & ${2.21}^{+0.14}_{-0.14} $ & ${2.21}^{+0.13}_{-0.13} $     \vspace{4px} \\ 
         &  & candidates & $-$ & $-$ & ${2.47}^{+0.10}_{-0.10} $ & ${2.02}^{+0.19}_{-0.34} $ & ${2.36}^{+0.11}_{-0.11} $ & ${2.32}^{+0.10}_{-0.10} $   \vspace{10px}\\ 
        
        Einstein radius & $\theta_{\rm E, sat}\ ['']$ & full chain &  $-$ & $-$ & ${0.30}^{+0.17}_{-0.23} $ & ${0.17}^{+0.15}_{-0.12} $ & ${0.23}^{+0.15}_{-0.12} $ & ${0.18}^{+0.17}_{-0.12} $     \vspace{4px} \\ 
         &  & candidates & $-$ & $-$ & ${0.18}^{+0.05}_{-0.05} $ & ${0.13}^{+0.12}_{-0.09} $ & ${0.22}^{+0.08}_{-0.07} $ & ${0.24}^{+0.17}_{-0.16} $    \vspace{10px}\\ 
        
        core radius & $r_{\rm c, sat}\ ['']$ & full chain &  $-$ & $-$ & $-$ & ${0.37}^{+0.12}_{-0.12} $ & $-$ & ${0.19}^{+0.21}_{-0.13} $     \vspace{4px} \\ 
         &  & candidates &  $-$ & $-$ & $-$ & ${0.34}^{+0.19}_{-0.13} $ & $-$ & ${0.24}^{+0.16}_{-0.12} $  \vspace{10px}\\ 
        \midrule
        \multicolumn{2}{l}{external shear} & & & & & \\ \midrule
        \ \\
        strength  & $\gamma_{\rm ext}$ & full chain &  ${0.07}^{+0.02}_{-0.02} $ & ${0.06}^{+0.02}_{-0.02} $ & $-$ & $-$ & ${0.06}^{+0.02}_{-0.02} $ & ${0.05}^{+0.02}_{-0.02} $  \vspace{4px} \\ 
         &  & candidates &  ${0.08}^{+0.02}_{-0.02} $ & ${0.08}^{+0.01}_{-0.02} $ & $-$ & $-$ & ${0.07}^{+0.02}_{-0.02} $ & ${0.05}^{+0.02}_{-0.02} $  \vspace{10px}\\ 
         
        position angle  & $\phi_{\rm ext}$ [\degree] & full chain &  ${31}^{+10}_{-8} $ & ${37}^{+7}_{-8} $ & $-$ & $-$ & ${33}^{+6}_{-7} $ & ${25}^{+16}_{-10} $    \vspace{4px} \\ 
         &  & candidates &   ${29}^{+8}_{-5} $  & ${32}^{+4}_{-3} $ & $-$ & $-$ & ${29}^{+3}_{-4} $ & ${18}^{+12}_{-8} $  \vspace{10px}\\ 

        \ \\ \midrule

        \end{tabularx}

        \end{table}

\end{landscape}

\clearpage

\section{Corner plots for candidate model classes}

In this Appendix, we present constraints on a selection of mass and external shear parameters for each candidate model class. The distributions of candidates are plotted in red over the general distribution of the complete chain, which is colored in black.

\begin{figure*}[h]
\centering
        \includegraphics[scale=0.45]{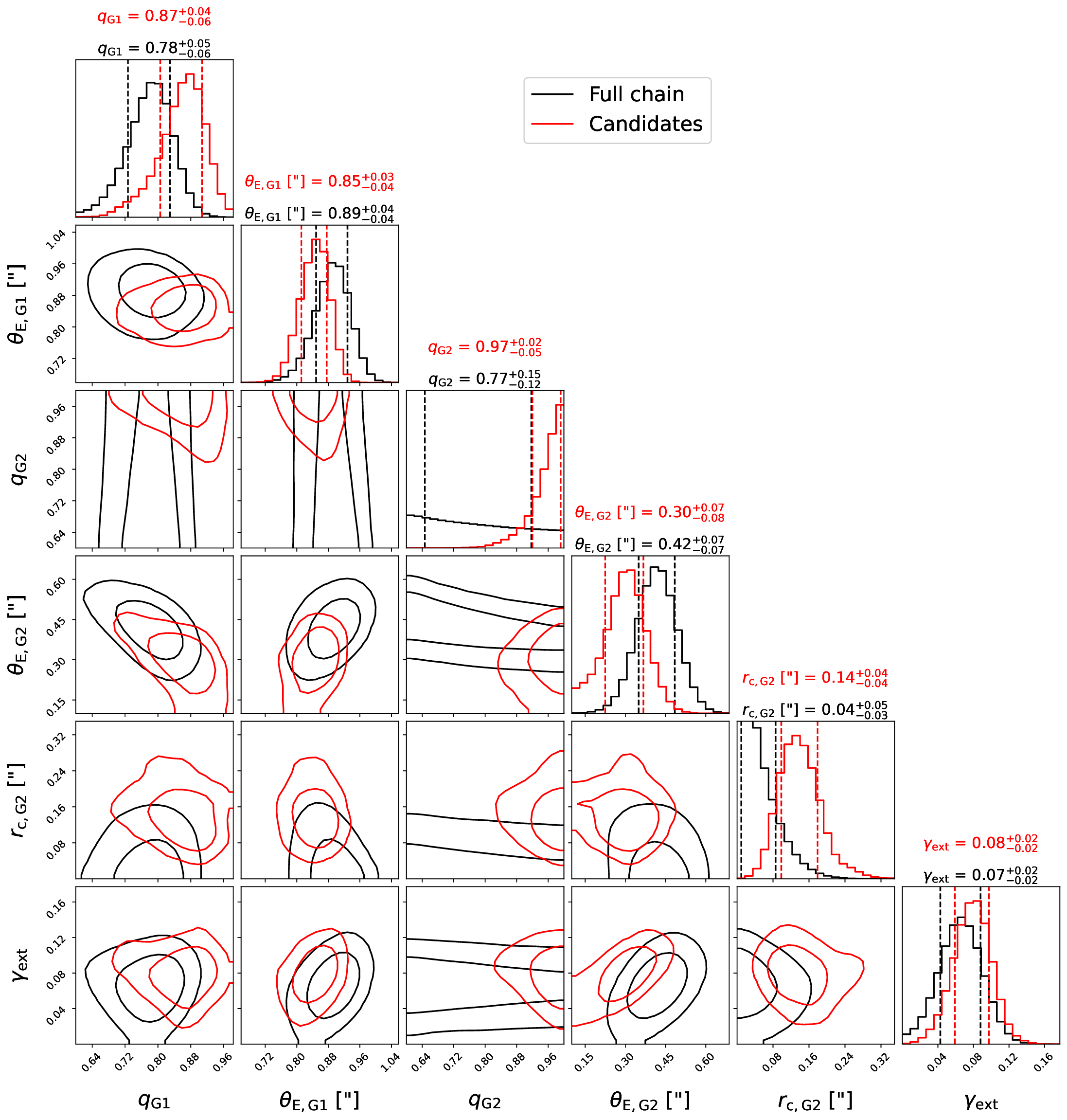}
        \caption{PL1 + cPL2 + $\gamma_{\rm ext}$: posterior distributions for several lens-mass parameters of the power-law profile (see Sect.~\ref{sec:modeling}) and the external shear strength. We show the distributions for candidate models (i.e., models that predict four observable quasar images) in red, and the distribution of the whole MCMC chain is plotted in black. The two contours show the 1$\sigma$ and 2$\sigma$ credible regions. The one-dimensional histograms show the marginalized posterior distribution for the selected mass parameters, and the vertical lines mark the 1$\sigma$ confidence intervals.}
        \label{fig:g2corner}
\end{figure*}

\begin{figure*}
\centering
        \includegraphics[scale=0.5]{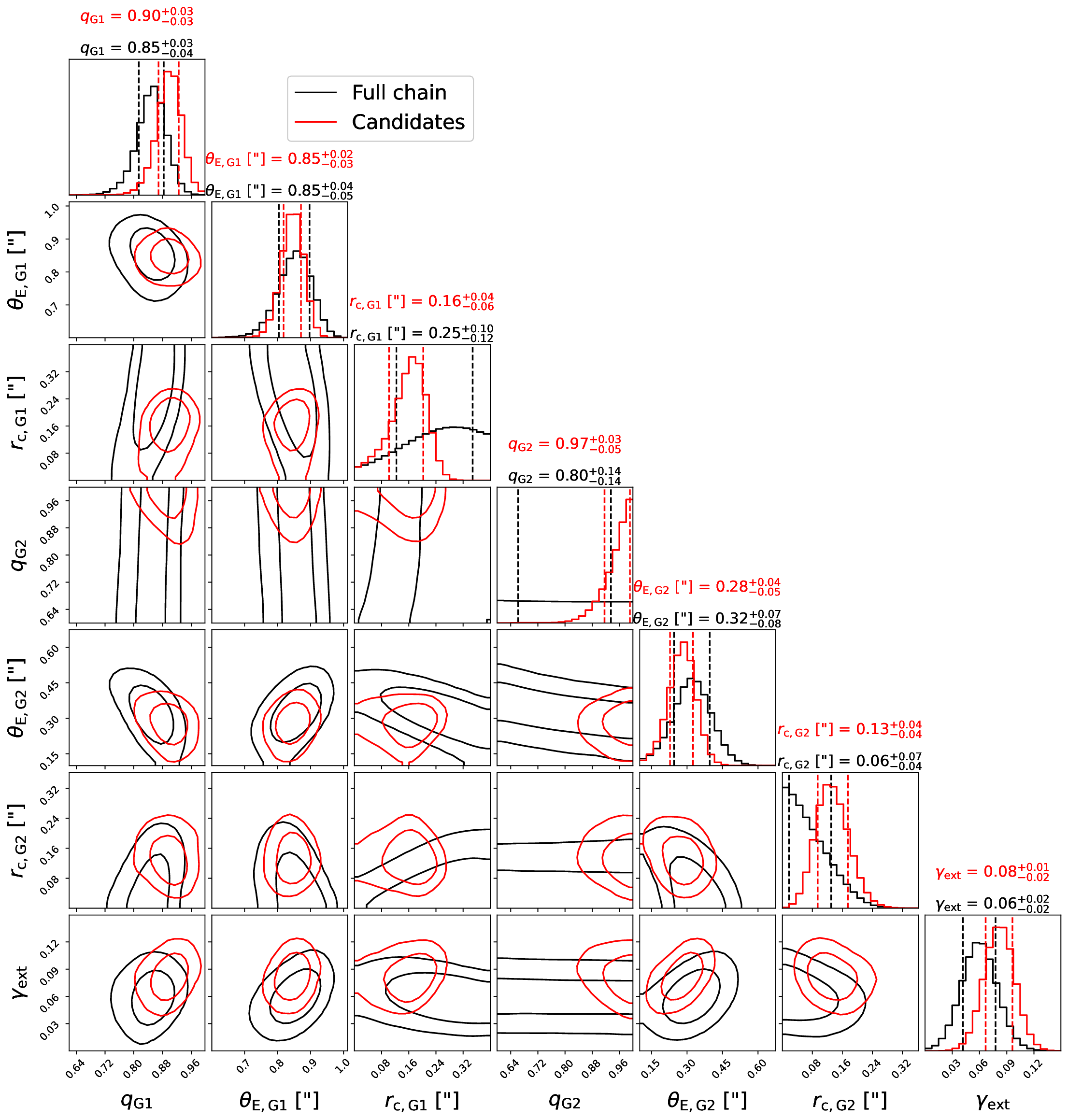}
        \caption{Posterior distributions for the cPL1 + cPL2 + $\gamma_{\rm ext}$ model class.}
        \label{fig:g1g2corner}
\end{figure*}

\begin{figure*}
\centering
        \includegraphics[scale=0.5]{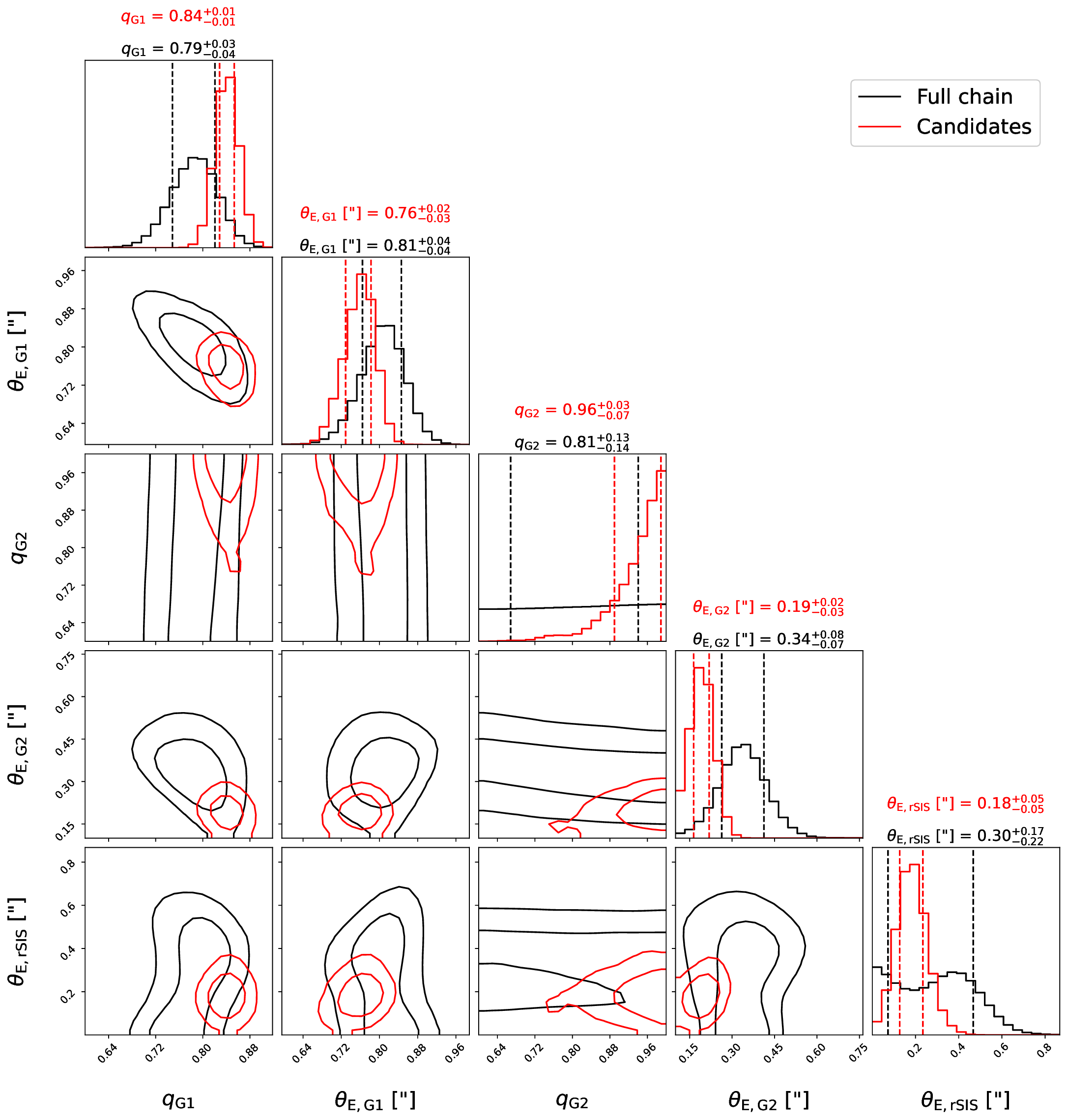}
        \caption{Posterior distributions for the PL1 + PL2 + rSIS model class.}
        \label{fig:rsiscorner}
\end{figure*}

\begin{figure*}
\centering
        \includegraphics[scale=0.5]{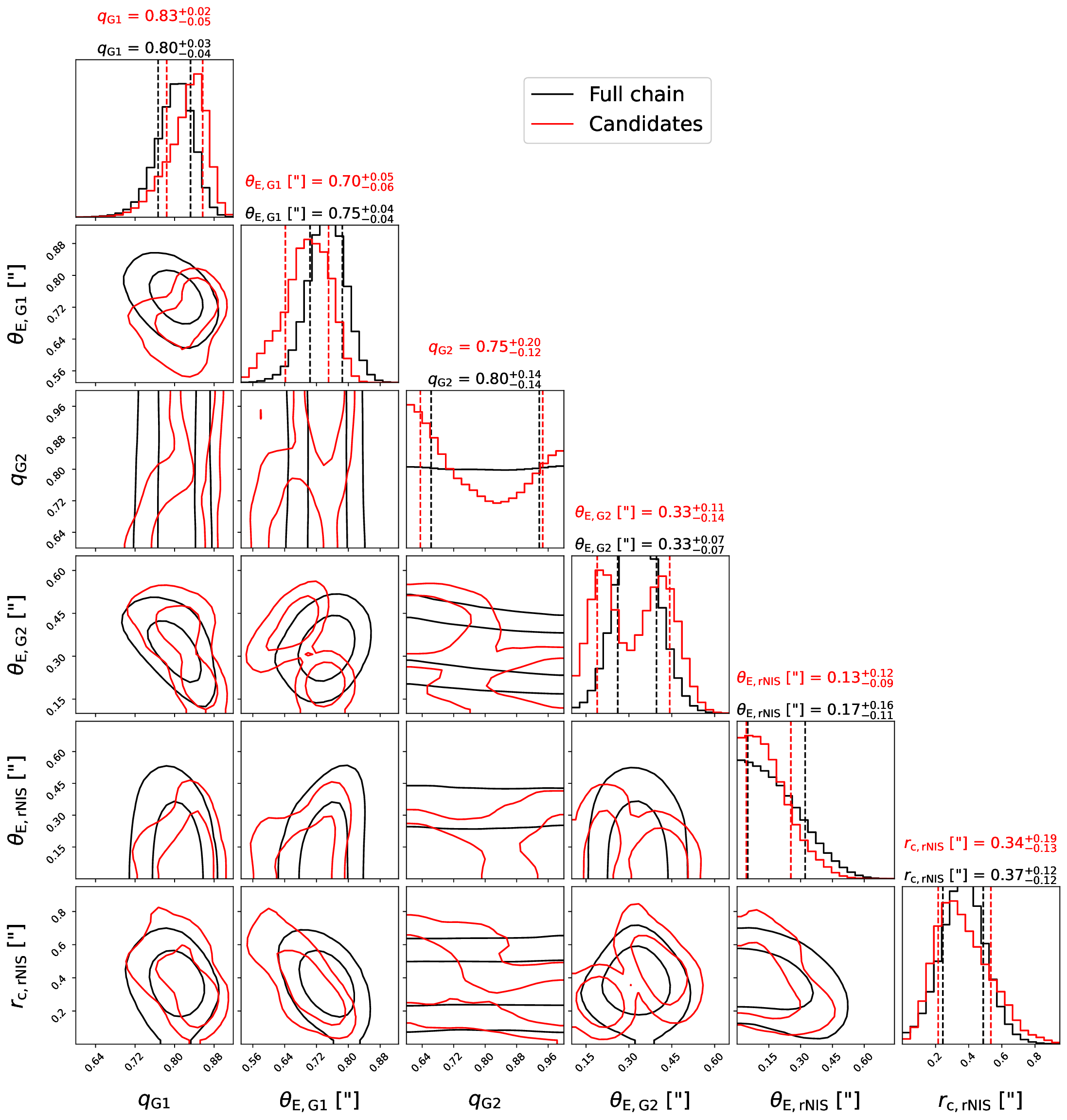}
        \caption{Posterior distributions for the PL1 + PL2 + rNIS model class.}
        \label{fig:rniscorner}
\end{figure*}

\begin{figure*}
\centering
        \includegraphics[scale=0.5]{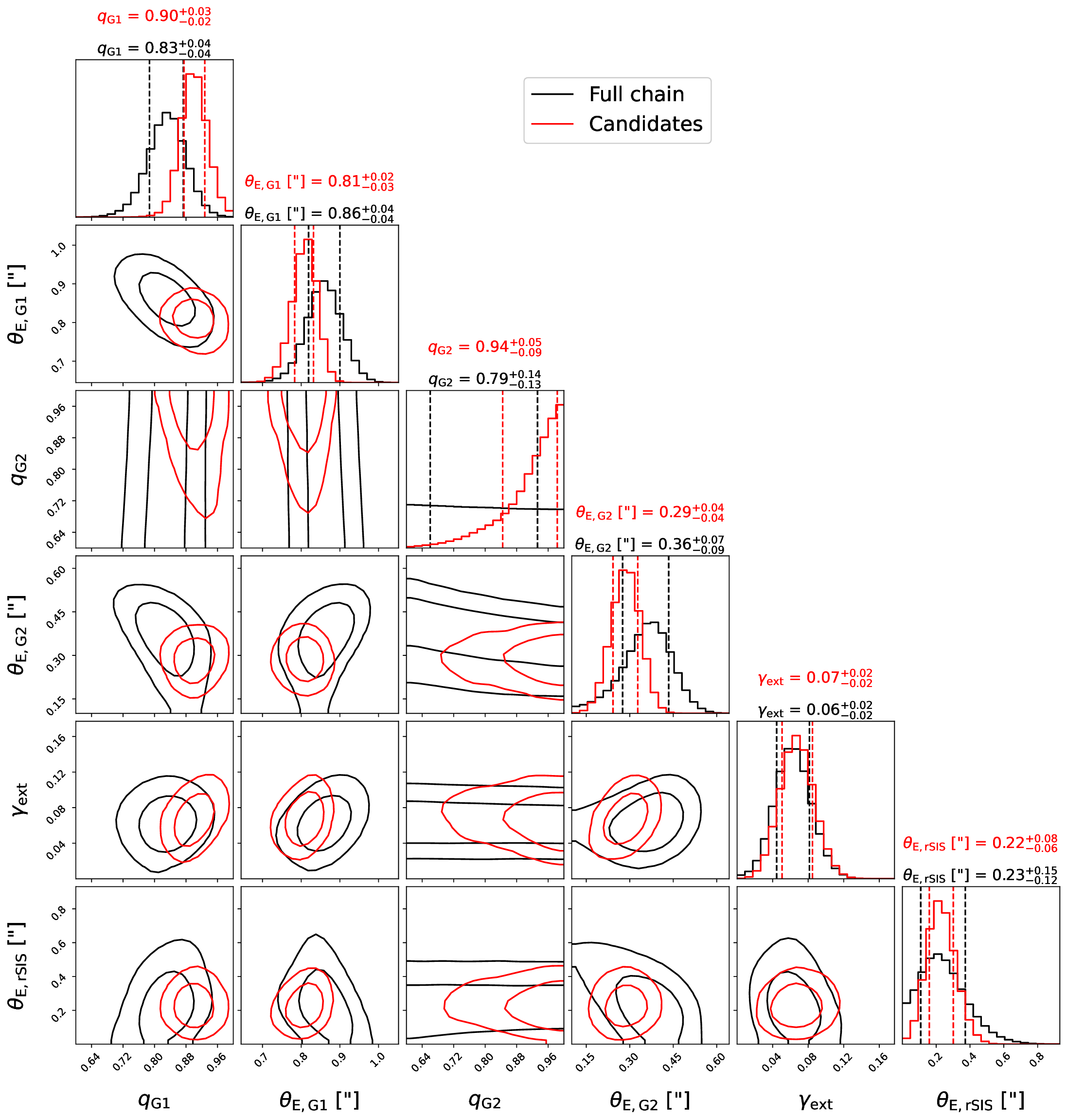}
        \caption{Posterior distributions for the PL1 + PL2 + rSIS + $\gamma_{\rm ext}$ model class.}
        \label{fig:rsis+scorner}
\end{figure*}

\begin{figure*}
\centering
        \includegraphics[scale=0.5]{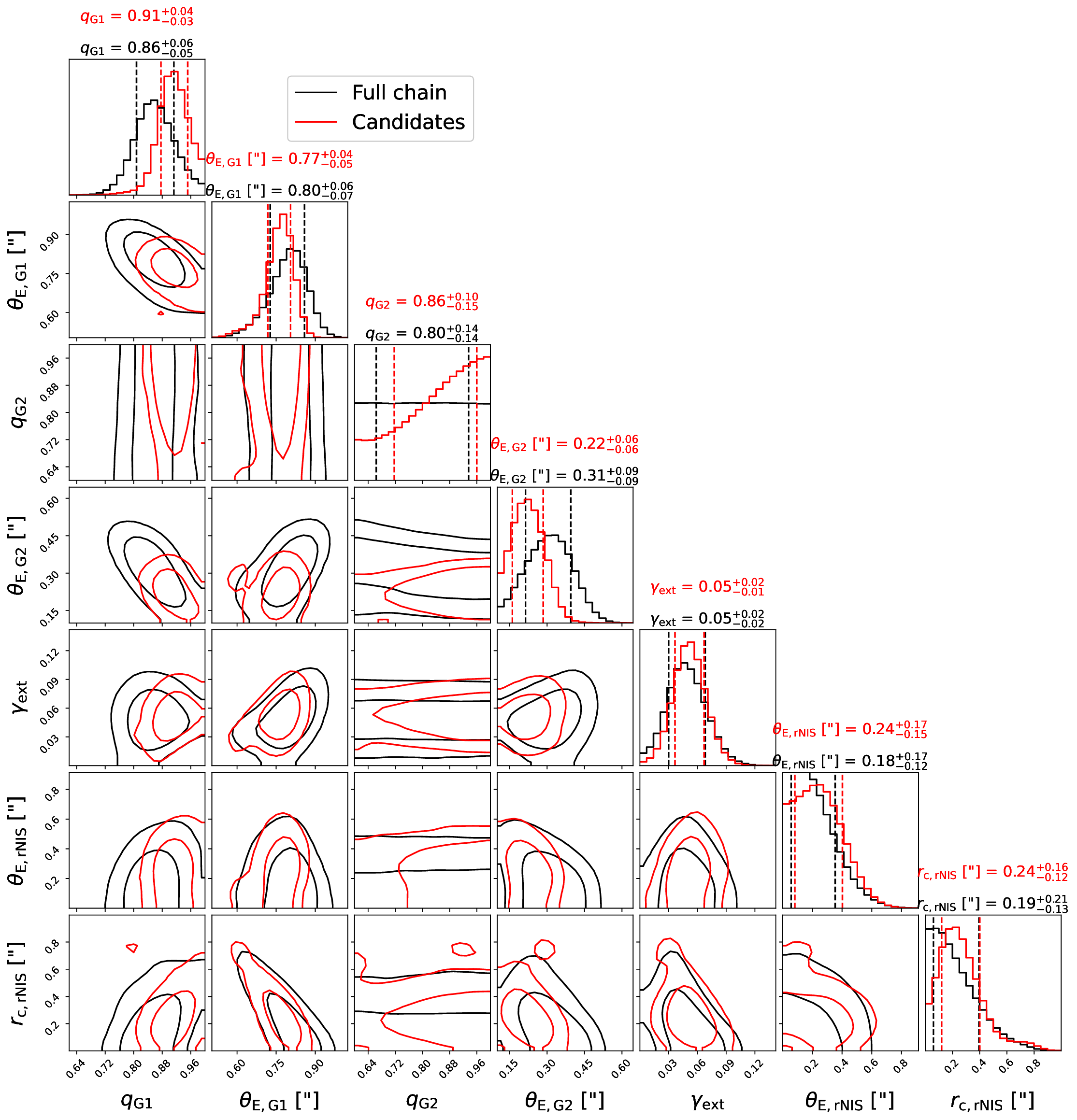}
        \caption{Posterior distributions for the PL1 + PL2 + rNIS + $\gamma_{\rm ext}$ model class.}
        \label{fig:rnis+scorner}
\end{figure*}
\end{appendix}

\end{document}